\documentclass[11pt,a4paper]{article}
\usepackage[truedimen,margin=30mm]{geometry} 

\usepackage{mathrsfs}
\usepackage{amssymb}
\usepackage{amsmath}
\usepackage{mathtools}
\usepackage{ascmac}
\usepackage{amsthm}
\usepackage{cmll}

\usepackage[pdftex]{graphicx}
\usepackage[pdftex]{color}

\usepackage{natbib}
\usepackage{setspace}
\usepackage{bm}
\usepackage{url}

\usepackage{color}

\usepackage{times}

\usepackage{titlesec}
\titleformat*{\section}{\large\bfseries}
\titleformat*{\subsection}{\it}

\newtheorem{thm}{Theorem}

\def\ep{{\varepsilon}}

\def\barbeta{{\bar{\beta}}}
\def\barmu{{\bar{\mu}}}

\def\tah{{\widehat{\tau}}}
\def\tat{{\widetilde{\tau}}}

\def\f{\mbox{\boldmath$f$}}

\def\mH{\mathcal{H}}

\def\mX{\mathcal{X}}
\def\mZ{\mathcal{Z}}

\def\seq#1#2{#1{:}#2}

\title{{\LARGE {\bf Bayesian Causal Synthesis for Meta-Inference on Heterogeneous Treatment Effects }}}

\date{}

\begin{document}

\maketitle
\doublespacing

\vspace{-1.5cm}
\begin{center}
{\large Shonosuke Sugasawa$^1$\footnote{Email: sugasawa@econ.keio.ac.jp}, K\={o}saku Takanashi$^2$, \\Kenichiro McAlinn$^3$, and Edoardo M. Airoldi$^3$}

\medskip

\medskip
\noindent
$^1$Faculty of Economics, Keio University\\
$^2$RIKEN Center for Advanced Intelligence Project\\
$^3$Department of Statistics, Operations, and Data Science, Fox School of Business, Temple University\\
\end{center}

\vspace{0.5cm}
\begin{center}
{\bf \large Abstract}
\end{center}
   
The estimation of heterogeneous treatment effects in the potential outcome setting is biased when there exists model misspecification or unobserved confounding.
As these biases are unobservable, what model to use when remains a critical open question.
In this paper, we propose a novel Bayesian methodology to mitigate misspecification and improve estimation via a synthesis of multiple causal estimates, which we call Bayesian causal synthesis.
Our development is built upon identifying a synthesis function that correctly specifies the heterogeneous treatment effect under no unobserved confounding, and achieves the irreducible bias under unobserved confounding.
We show that our proposed method results in consistent estimates of the heterogeneous treatment effect; either with no bias or with irreducible bias.
We provide a computational algorithm for fast posterior sampling.
Several benchmark simulations and an empirical study highlight the efficacy of the proposed approach compared to existing methodologies, providing improved point and density estimation of the heterogeneous treatment effect, even under unobserved confounding. 

\vspace{-0cm}


\newpage
\section{Introduction}\label{sec:Intro}

One of the central problems of statistics is assessing, comparing, and mitigating model misspecification.
In a predictive context, this is fairly straightforward: the accuracy of a model is measured by how well it predicts future, out-of-sample data.
By having {\it direct} feedback (i.e., mean squared error, predictive likelihood, etc.), we can measure the efficacy of any proposed method-- either model selection or combination-- through its predictive performance.
In a causal inference context, however, we have a starkly different picture.
As all potential outcomes of a unit can never be observed (e.g. we can never observe what would have happened if a patient did not take a drug, if they did), there is no direct way to measure how well a method is predicting, or in the causal inference jargon {\it imputing}, the unobserved potential outcome.
The causal inference literature has traditionally dealt with this problem through design of experiments techniques, both in randomized controlled trials and in observational studies, assuring that, under some assumptions, the target treatment effect can be estimated with some guarantees \citep{rubin1974estimating,rubin2008forobjective,imbens2015causal,ding2023first}.

The broadening of applications in causal inference, especially at the intersection of economics and machine learning, often times in an observational study setting, has led to developments of many model-based methods that allow for complex analyses of the target treatment effects, particularly in the Bayesian literature \citep{gustafson2006curious,hill2011bayesian,heckman2014treatment,li2014bayesian,hahn2018regularization,roy2018bayesian,sugasawa2019estimating}. However, in these applications, assumptions are complex and harder to meet or justify, and the model-based nature of the methodology being proposed leaves the door open to issues of model misspecification. 
One particularly relevant and prominent example of this is the problem of estimating heterogeneous treatment effects \citep[HTE:][]{green2012modeling,athey2016recursive,taddy2016nonparametric,henderson2016bayesian,powers2018some,kunzel2019metalearners,syrgkanis2019machine}, where modeling assumptions are most often key to assessing heterogeneity.

We contribute to this literature by proposing a formal Bayesian approach to synthesize multiple HTE estimates, which we call Bayesian causal synthesis (BCS).
Rather than finding the best method/model/specification, our approach leverages the discrepancies between different estimates, learns from their biases and dependencies, and improves HTE estimation through a synthesis of information.
By not relying on a single estimate, we are  conducting what we call meta-inference, a way to infer {\it above} the individual estimates through their synthesis, which not only robustifies estimation, but also expands the model space to better approximate the underlying heterogeneity.
This process is analogous to meta-analysis, though we use meta-inference to denote a wider class of problems that involve model misspecification and data uncertainty.

Three problems make ensemble methods for causal inference particularly difficult.
The first is that there is no clear target to minimize, and it is not obvious whether modeling the observed outcome leads to improved estimation of the HTE.
The second, which relates to the first, is that there exists performance heterogeneity in estimating HTEs.
In other words, one estimator can be good at estimating the treatment effect on one group but poor in another, and vice versa.
Thus, minimizing some criteria in a sweeping manner (as is done in standard model averaging) cannot capture this performance heterogeneity.
The third is that these estimates are highly dependent, due to their shared usage of data, modeling structure, etc., making simple ensembling (e.g. linear averaging, such as stacking)-- which implicitly assumes that each estimate is independent-- inadequate.
Existing ensemble methods cannot take into account, or mitigate, these issues, due to their construction and assumptions.

Our BCS approach mitigates all three issues through the usage of the Bayesian predictive synthesis (BPS) framework \citep[][]{mcalinn2019dynamic}.
While the BPS framework has been shown to be superior to other, benchmark and state-of-the-art, ensemble methods in predictive tasks \citep[e.g.][]{mcalinn2020multivariate,cabel2022spatially}, the extension to causal inference has not been done due to the aforementioned problems (particularly the problem of linking prediction to causal estimation).
For BCS, we develop a synthesis function we identify to have desirable theoretical properties; correct specification under no unobserved confounding and irreducible bias under unobserved confounding.
The synthesis function we propose has two components: the first is the prognostic term that includes the propensity score, and the second synthesizes the HTE estimates-- treated as latent factors-- through a Gaussian process.
We show that this approach achieves consistency with regard to the HTE estimate, through proving predictive consistency of the treated and control.
This result is crucial, as this links the connection between the predictive aspect of the data and the accuracy of its causal estimate.
As such, this result, not only provides theoretical guarantees regarding the estimate, it provides methodological justification for the approach.

Comparisons using several simulation data and competition data show that our approach improves upon existing approaches, in terms of both mean and coverage probabilities.
In simulated data with unobserved confounding, we show that BCS outperforms statistical and machine learning based HTE methods, under multiple data generating mechanisms and dimensions.
Even with unobserved confounding causing bias, the coverage probability of BCS is within $\pm$4\% of the nominal coverage; a stark difference from the best competing strategy with a coverage of 84.4\%.
Using the dataset from the Atlantic causal inference conference data analysis challenge (2017), we outperform competing strategies in the most difficult scenarios, achieving near ideal coverage probability (within approximately  95$\pm$2\% for all scenarios considered).
Finally, we demonstrate the flexibility of our approach in a re-analysis of data from an observational study on the effect of smoking on medical expenditures.

\subsection{Related work}\label{sec:related}

The literature regarding the problem of estimating heterogeneous treatment effects is burgeoning. 
Recent advances in machine learning methods for causal inference have led to new estimators for heterogeneous treatment effects, including  causal forest \citep[CF:][]{wager2018estimation,athey2019estimating,athey2019generalized} and causal BART \citep[BCF:][]{hahn2020bayesian,linero2022mediation}, which have both received considerable attention.
The appeal for these methods is, in part, their flexibility, achieved through sum-of-regression-trees.
However, despite their flexibility, these methods can produce widely different estimates, as they are sensitive to model specification, prior specifications, etc., particularly when the number of observations is small in any given group.
In practice, this is problematic for the user, as there is no {\it right} method or set of specifications that works well across most situations. The theory underlying these methods is also limited to toy settings.

These issues have been reported in the literature, as evidenced by large-scale simulation studies and causal inference competitions that clearly show that no single method performs the best in every problem setting \citep{dorie2017aciccomp2016,wendling2018comparing,carvalho2019assessing,dorie2019automated,hahn2019atlantic,mcconnell2019estimating}.
Yet, there is no straightforward way to deal with model uncertainty in this context, due to the aforementioned problem of unobserved potential outcomes, and standard model selection techniques (such as information criteria or cross validation) are not adequate.

The idea of using multiple estimates to improve estimation-- often coined as ensemble methods-- is not new \citep[in fact, the idea can be traced back to, at least,][]{galton1907vox}.
In the predictive literature, ensemble methods have been used extensively to improve predictions by reducing model uncertainty \citep{bates1969combination,hall2007combining,wang2022forecast}.
As the task typically involves minimizing some predictive criteria, developing ensemble methods (including estimating ensemble weights) is straightforward; at the very least, the efficacy can be directly assessed through predictive accuracy, even in real world data.
This, again, is not the case for causal inference.
Nonetheless, many causal inference methods-- implicitly or explicitly-- rely on ensembling ideas.
For example, CF and BCF both use a sum of trees, which is an averaging of individual trees (in the case of CF, this is an equal weight average), and \cite{zigler2014uncertainty} uses model averaging to model the propensity score.
Related to this paper, \cite{han2022ensemble} propose a stacking approach to ensemble individual treatment effects.

\subsection{Organization and contributions of this paper}\label{sec:org}

In Section~\ref{sec:BCS}, we introduce the problem statement and notation, and propose the BCS framework.
Our proposed synthesis function is based on theoretically identifying a model that correctly specifies the HTE under no unobserved confounders.
We then show that this synthesis function achieves the irreducible bias under unobserved confounding.
We then show that our proposed BCS produces consistent estimates of the true HTE or to the irreducible bias, depending on the existence of unobserved confounding.
Section~\ref{sec:comp} describes the exact implementation and posterior computation algorithm for analysis.
We conduct two sets of simulation studies in Section~\ref{sec:sim}, where we compare our proposed BCS to standard and state-of-the-art estimation methods in the statistical and machine learning literature.
The first simulation study considers a total of four scenarios with varying dimensions under unobserved confounding.
The second simulation study considers eight scenarios from a causal inference competition.
Section~\ref{sec:app} presents an empirical application to medical expenditure data to highlight the usability and interpretability of BCS.
Section~\ref{sec:summ} concludes the paper with summary comments.
R code implementing the proposed method is available at GitHub repository (\url{https://github.com/sshonosuke/BCS}).

\section{Bayesian Causal Synthesis \label{sec:BCS}}

\subsection{Problem setting and notation}

Let $Z\in \{0,1\}$ be the binary treatment indicator, with $Z=1$ indicating being treated and $Z=0$ indicating the control. 
Let $Y{(Z)}$ denote the potential outcome indicating the response of a subject if the subject receives treatment $Z$. The following methods can be generally adapted to various types of outcomes.
In practice, we can only observe $Y=Y{(Z)}$ for each subject, that is, each patient has potential outcomes $Y{(1)}$ and $Y{(0)}$, but only one of them can be observed as $Y$. 
We assume that $X$, a $q$-dimensional covariate vector, exists. 
Hence, the data consists of sets of these variables; $\{(y_i, x_i, Z_i),\ i=1,\ldots,n\}$.

Let $p_{x}$ denote the probability density from which the covariates,
$\boldsymbol{x}$, follow. Let $H_{x}$ be the Hilbert space constructed
from the square-integrable functions regarding $p_{x}$, and define
the inner product as 
\[
\left\langle g,h\right\rangle _{H_{x}}=\mathbb{E}_{x}\left[g\left(\boldsymbol{x}\right)h\left(\boldsymbol{x}\right)\right]=\int g\left(\boldsymbol{x}\right)h\left(\boldsymbol{x}\right)p_{x}\left(\boldsymbol{x}\right)d\boldsymbol{x},\ g,h\in H_{x}.
\]
Let the outcome variable, $Y\left(Z,X\right)$,
be a Gaussian random variable on $H_{x}$, such that $Y\left(Z,X\right)=ZY\left(Z=1,X\right)+\left(1-Z\right)Y\left(Z=0,X\right)$
and $\mu_{Y}$ be the probability measure on $H_{x}$. Here, we write
$Y\left(Z=1,X\right),Y\left(Z=0,X\right)$ to indicate whether
the outcome variable is treated, $Z=1$, or not, $Z=0$. The probability
measures for $Y\left(Z=1,X\right),Y\left(Z=0,X\right)$ on
$H_{x}$ are denoted as $\mu_{Y({1})},\mu_{Y({0})}$, respectively.
If we fix one $\boldsymbol{x}$, we have 
\begin{equation}
\begin{alignedat}{1}Y\left(Z=1,\boldsymbol{x}\right) & \sim\mathcal{N}\left(F\left(Z=1,\boldsymbol{x}\right),\sigma\left(Z=1,\boldsymbol{x}\right)\right),\\
Y\left(Z=0,\boldsymbol{x}\right) & \sim\mathcal{N}\left(F\left(Z=0,\boldsymbol{x}\right),\sigma\left(Z=0,\boldsymbol{x}\right)\right).
\end{alignedat}
\label{eq:DGP_field-Gauss}
\end{equation}
Our main goal is to estimate the heterogeneous treatment effect (HTE): the difference between the response $Y$, under $Z=1$ and $Z=0$, in hypothetical worlds, averaged across subpopulations defined by covariates, $X$. 
This class of counterfactual estimands can be formalized in the potential outcome framework \citep{rubin2005causal,imbens2015causal}.
We make the stable unit treatment value assumption (SUTVA) throughout  \citep[excluding interference between units and multiple versions of treatment;][]{imbens2015causal}.
Thus, we observe the potential outcome that corresponds to the realized treatment, $y_i=Z_iY_i{(1)}+(1-Z_i)Y_i{(0)}$.

As noted in Section~\ref{sec:Intro}, there are a variety of methods to estimate $\tau(X)$ from the observed data, and our goal is to develop a methodology to synthesize multiple estimates to make meta-inference on $\tau(X)$.
Specifically, let $\tah_j(X) \ (j=1,\ldots,J)$ be an estimate of $\tau(X)$ by the $j$-th method and $s_j(X)$ be the associated standard error of $\tah_j(X)$. 
Then, the approximated distribution of the treatment effect, which we denote as $h_j(X)$, can be defined as a normal distribution with mean $\tah_j(X)$ and variance $s_j^2(X)$.  

\subsection{Strong ignorability and unobserved confounding}

A common assumption made in this context is the strong ignorability condition, which is defined as follows.
Let $\boldsymbol{x}$ be the full set of covariates that characterizes the DGP in eq.~\eqref{eq:DGP_field-Gauss}.
The strong ignorability condition states that $(Y_{i}{(1)},Y_{i}{(0)})$
and $Z_{i}$ are independent given $\boldsymbol{x}_{i}$: 
\[
\left.\left(Y_{i}\left(1\right),Y_{i}\left(0\right)\right)\Perp Z_{i}\right|\boldsymbol{x}_{i},
\]
 and also that $0<P\left(\left.Z_{i}=1\right|\boldsymbol{x}_{i}\right)<1$
for $i=1,\ldots,n$.

In many contexts, this full set of covariates will not be available.
Then, we can partition $\boldsymbol{x}$ as $\boldsymbol{x}=\left[\boldsymbol{x}^{\textrm{obs}\top},\boldsymbol{x}^{\textrm{mis}\top}\right]^{\top}$,
where $\boldsymbol{x}^{\textrm{obs}}$ is the subvector of observable covariates, and $\boldsymbol{x}^{\textrm{mis}}$ is the subvector of missing covariates.
If the strong ignorability condition holds with only the observed confounders, i.e.,
\[
\left(Y_{i}\left(1\right),Y_{i}\left(0\right)\right)
\Perp
Z_{i} \mid \boldsymbol{x}_{i}^{\textrm{obs}},
\]
then $\boldsymbol{x}^{\textrm{mis}}=\varnothing$, $\boldsymbol{x}^{\textrm{obs}}=\boldsymbol{x}$ and $E[Y_{i}{(Z_{i})}|x_{i}]=E[y_{i}|Z_{i},x_{i}]$. The
HTE can be expressed as 
\begin{alignat*}{1}
\tau\left(\boldsymbol{x}_{i}\right) & \equiv\mathbb{E}\left[\left.Y_{i}\left(Z_{i}=1,\boldsymbol{x}_{i}\right)\right|\boldsymbol{x}_{i}\right]-\mathbb{E}\left[\left.Y_{i}\left(Z_{i}=0,\boldsymbol{x}_{i}\right)\right|\boldsymbol{x}_{i}\right].\\
 & =\mathbb{E}\left[\left.Y_{i}\left(Z_{i}=1,\boldsymbol{x}_{i}\right)\right|Z_{i}=1,\boldsymbol{x}_{i}\right]-\mathbb{E}\left[\left.Y_{i}\left(Z_{i}=0,\boldsymbol{x}_{i}\right)\right|Z_{i}=0,\boldsymbol{x}_{i}\right],
\end{alignat*}
and there are no unobserved confounders.
If, however, $\boldsymbol{x}^{\textrm{mis}}$ affects both the outcome, $Y$, and treatment assignment, $Z$, the strong ignorability condition will not hold due to unobserved confounding.
Specifically, given strong ignorabililty under $\boldsymbol{x}^{\textrm{obs}},\boldsymbol{x}^{\textrm{mis}}$,
\[
\left.\left(Y_{i}\left(1\right),Y_{i}\left(0\right)\right)\Perp Z_{i}\right|\boldsymbol{x}_{i}^{\textrm{obs}},\boldsymbol{x}_{i}^{\textrm{mis}},
\]
if there exists unobserved confounding, the strong ignorability condition will not hold with only $\boldsymbol{x}^{\textrm{obs}}$:
\[
\left(Y_{i}\left(1\right),Y_{i}\left(0\right)\right)
\not\Perp
Z_{i} \mid \boldsymbol{x}_{i}^{\textrm{obs}}.
\]
Then, we have
\[
p\left(\left.Y\right|Z,\boldsymbol{x}^{\textrm{obs}}\right)\neq p\left(\left.Y\right|\boldsymbol{x}^{\textrm{obs}}\right).
\]
Under this scenario, the estimated HTE will be biased due to the effect of $\boldsymbol{x}^{\textrm{mis}}$ not being measurable.
Since these are fundamentally unobserved, this bias is irreducible.

\subsection{Synthesis model for multiple treatment effect estimates \label{sec:BCSintro}}

We will first consider the case where there are no unobserved confounding, then consider the case where there are, in Section~\ref{sec:conf}.

Let $f_j(X)$ be a random variable following the $j$-th approximated distribution of $\tau(X)$,  $h_j(f(X))$, which comprises the set, $\mH=\{h_1(X),\ldots,h_J(X)\}.$
Further, denote the estimator predictive values as $\left\{ f_{j}\left(Z=1,X\right),f_{j}\left(Z=0,X\right)\right\} _{j=1{:}J}$.
Then, $\left\{ f_{j}\left(Z=1,X\right),f_{j}\left(Z=0,X\right)\right\} _{j=1{:}J}$
are each Gaussian random variables taking values on $H_{x}$, and
let $\left\{ \mu_{f_{j}({1})},\mu_{f_{j}({0})}\right\} _{j=1{:}J}$
denote probability measures on $H_{x}$. Note that 
\[
f_{j}\left(X\right)=f_{j}\left(Z=1,X\right)-f_{j}\left(Z=0,X\right).
\]
Then, in the BPS framework, the set, $\mH$, is synthesized via Bayesian updating with the posterior of the form,
\begin{equation}
p(Y|\mH ,X)=\int_{\f(X)}\alpha(Y|\f(X))\prod_{j=\seq1J}h_{j}(X)d{\f(X)},\label{eq:theorem1}
\end{equation}
where $\f(X)=(f_{1}(X),\ldots,f_{J}(X))^{\top}$ is a $J$-dimensional latent vector and $\alpha(Y|\f(X))$ is a conditional probability density function for $Y$ given
$\f(X)$, called the synthesis function.
Note that eq.~\eqref{eq:theorem1} is only a coherent Bayesian posterior if it satisfies the consistency condition \citep[see,][]{genest1985modeling,west1992modelling1,west1992modelling2,mcalinn2019dynamic}.
In short, the consistency condition states that, given the decision maker's prior, $p(Y)$, and their prior of what each estimator will produce before observing it, $m(\f(X))$, the priors have to be consistent: $p(Y)=\int_{\f(X)}\alpha(y|\f(X))m(\f(X))d\f(X)$, for eq.~\eqref{eq:theorem1} to be a coherent Bayesian posterior.

While the theory of the BPS framework does not specify what the synthesis function is, we theoretically derive a synthesis function that has desirable theoretical properties, e.g. consistency, regarding the estimation of the HTE.
Denote $\mu_{Y,\boldsymbol{f}}$ as joint probability measure of
$(Y(X),\{ f_{j}(Z=1,X)\} _{j=1{:}J},\{ f_{j}(Z=0,X)\}_{j=1{:}J})$.
The HTE conditional on $\boldsymbol{x}$ is defined as $\mathbb{E}[Y(Z=1,X)-Y(Z=0,X)\mid \boldsymbol{x}]$.
The goal is to construct the theoretically best estimator for the HTE given $\boldsymbol{x}$
by predicting $Y\left(Z=1,X\right),Y\left(Z=0,X\right)$ using
the estimator predictive values, $\{ f_{j}(Z=1,X),f_{j}(Z=0,X)\}_{j=1{:}J}$.
Consider at some point we have $n$ worth of observed data and estimator
predictive values, $\left\{ \left(y({Z_{s}}),\boldsymbol{x}_{s}\right)\right\} _{1\leq s\leq n}$,
$\left\{ f_{j}\left(Z=1,\boldsymbol{x}_{s}\right),f_{j}\left(Z=0,\boldsymbol{x}_{s}\right)\right\}_{1\leq s\leq n, \ j=1{:}J}$,
and denote the $\sigma$-algebra generated from this set, $\mathcal{G}$.
From the $\mathcal{G}$-conditional expectation, $Y\left(X\right)$
can be decomposed into 
\begin{equation}
Y\left(Z,X\right)=\mathbb{E}_{\mu_{Y,\boldsymbol{f}}}\left[\left.Y\left(Z,X\right)\right|\mathcal{G}\right]+\varepsilon\left(Z,X\right),\ \left(\varepsilon\left(Z,X\right)\textrm{ is independent of }\mathcal{G}\right).\label{eq:SVCM_1}
\end{equation}
If we can identify, $\mathbb{E}_{\mu_{Y,\boldsymbol{f}}}\left[\left.Y\left(Z,X\right)\right|\mathcal{G},Z\right]$,
from
\[
\mathbb{E}\left[\left.Y\left(Z=1,X\right)-Y\left(Z=0,X\right)\right|\boldsymbol{x}_{n+1}\right]=\mathbb{E}\left[\left.\mathbb{E}_{\mu_{Y,\boldsymbol{f}}}\left[\left.Y\left(Z=1,X\right)-Y\left(Z=0,X\right)\right|\mathcal{G}\right]\right|\boldsymbol{x}_{n+1}\right],
\]
we can estimate the HTE as $\mathbb{E}\left[\left.\mathbb{E}_{\mu_{Y,\boldsymbol{f}}}\left[\left.Y\left(Z=1,X\right)-Y\left(Z=0,X\right)\right|\mathcal{G}\right]\right|\boldsymbol{x}_{n+1}\right]$.
This, as it turns out, can be expressed using a varying coefficient
model. 

\begin{thm}\label{thm:VCM} Let the estimator predictive values, $f_{j,n+1}\left(Z=1,\boldsymbol{x}_{n+1}\right),f_{j,n+1}\left(Z=0,\boldsymbol{x}_{n+1}\right)$,
be a Gaussian field with $\boldsymbol{x}_{n+1}$ as parameters. Here,
there exists a $\mathcal{G}$-measurable function, $\beta_{0}\left(Z=1,X\right)$, $\left\{ \beta_{j}\left(Z=1,X\right)\right\} _{j=1{:}J}$, 
$\beta_{0}\left(Z=0,X\right)$, $\left\{ \beta_{j}\left(Z=0,X\right)\right\} _{j=1{:}J}$,
and we have the expression, 
\begin{equation}
\begin{alignedat}{1}Y\left(Z=1,\boldsymbol{x}_{n+1}\right) & =\beta_{0}\left(Z=1,\boldsymbol{x}_{n+1}\right)+\sum_{j=1}^{J}\beta_{j}\left(Z=1,\boldsymbol{x}_{n+1}\right)f_{j}\left(Z=1,\boldsymbol{x}_{n+1}\right)+\varepsilon\left(Z=1,\boldsymbol{x}_{n+1}\right),\\
 & \mathbb{E}_{\mu_{Y,f}}\left[\left.\varepsilon\left(Z=1,\boldsymbol{x}_{n+1}\right)\right|\mathcal{G}\right]=0,\ \left(^{\forall}\boldsymbol{x}_{n+1}\right).\\
Y\left(Z=0,\boldsymbol{x}_{n+1}\right) & =\beta_{0}\left(Z=0,\boldsymbol{x}_{n+1}\right)+\sum_{j=1}^{J}\beta_{j}\left(Z=0,\boldsymbol{x}_{n+1}\right)f_{j}\left(Z=0,\boldsymbol{x}_{n+1}\right)+\varepsilon\left(Z=0,\boldsymbol{x}_{n+1}\right),\\
 & \mathbb{E}_{\mu_{Y,f}}\left[\left.\varepsilon\left(Z=0,\boldsymbol{x}_{n+1}\right)\right|\mathcal{G}\right]=0,\ \left(^{\forall}\boldsymbol{x}_{n+1}\right).
\end{alignedat}
\label{eq:SVCM_2}
\end{equation}
\end{thm}
The proof of Theorem~\ref{thm:VCM} is in the Supplementary Material (Section~\ref{supp:th1}).
This result shows that the HTE residual has an expectation of zero when a varying coefficient model is used as a synthesis function.
Thus, the HTE is correctly specified under this synthesis function.

This result, eq.~\eqref{eq:SVCM_2}, states that a nonparametric function, $\beta_{0}\left(Z,X\right)$,
$\left\{ \beta_{j}\left(Z,X\right)\right\} _{j=1{:}J}$, that synthesizes multiple estimates is essentially required to represent the DGP in eq.~\eqref{eq:DGP_field-Gauss}.
This also implies that if the intercept, $\beta_{0}\left(Z,X\right)$, correctly captures eq.~\eqref{eq:DGP_field-Gauss}, there is no need to synthesize multiple estimates.
However, since the intercept is a fully nonparametric function, it suffer under the curse of dimensionality.
When conditioning on the estimators, $\left\{ f_{j,n+1}\left(Z,\boldsymbol{x}_{n+1}\right)\right\} _{j=1{:}J}$, depending on how well they capture eq.~\eqref{eq:DGP_field-Gauss}, the coefficient function, $\left\{ \beta_{j}\left(Z,X\right)\right\} _{j=1{:}J}$, becomes smooth, with the intercept becoming zero or some constant near it.
Thus, the synthesis of estimators reduces the parameter complexity.
Under unobserved confounders, there are further merits in taking this synthesis approach (as discussed in the next subsection).

Given the results in Theorem~\ref{thm:VCM}, we propose the following synthesis model as the conditional specification of $\alpha(Y|\f(X))$, which is equivalent to eq.~(\ref{eq:SVCM_2}) in terms of its conditional expectation: 
\begin{equation}\label{BPS}
Y=\mu(X, \pi)+Z\left\{\beta_0(X)+\sum_{j=1}^J \beta_j(X)f_j(X)\right\}+\ep, \ \ \  \ep\sim N(0, \sigma^2),
\end{equation}
where $\mu(X,\pi)$ is an unknown function of $X$ and propensity score, $\pi=P(Z=1|X)$, $f_j(X)$ is the $j$-th latent variable having distribution $h_j(X)$, and $\beta_j(X)$ is the weight for the $j$-th estimator. 
The HTE under the synthesis model (eq.~\ref{BPS}) is given by 
\begin{equation}\label{eq:tau}
\tat(X)\equiv 
E[Y|Z=1,X]-E[Y|Z=0,X]=\beta_0(X)+\sum_{j=1}^J \beta_j(X)f_j(X),
\end{equation}
and $\mu(X, \pi)$ corresponds to the prognostic term.

\subsection{Best approximate model under unobserved confounding\label{sec:conf}}
While the result in Theorem~\ref{thm:VCM} holds when there are no
unobserved confounding, this assumption does not hold for most real world data.
We will now consider the performance of the synthesis function in
eq.~(\ref{BPS}) under unobserved confounding.
Specifically, we will show that our proposed method produces the best approximation of the true HTE, up to the irreducible bias from unobserved confounding, in terms of mean squared error (best approximation is defined in eq.~\ref{def:bam}).

As mentioned previously, under unobserved confounding, the true HTE,
\[
\tau\left(\boldsymbol{x}^{\textrm{obs}}\right)=\mathbb{E}\left[\left.Y\left(Z=1,\boldsymbol{x}\right)-Y\left(Z=0,\boldsymbol{x}\right)\right|\boldsymbol{x}^{\textrm{obs}}\right],
\]
is impossible to estimate because the omitted variable bias is unknown.
This bias can be characterized as follows. For each of the outcome,
$\mathbb{E}\left[\left.Y_{i}\left(Z_{i}=1,\boldsymbol{x}_{i}\right)\right|\boldsymbol{x}_{i}\right]$
and $\mathbb{E}\left[\left.Y_{i}\left(Z_{i}=0,\boldsymbol{x}_{i}\right)\right|\boldsymbol{x}_{i}\right]$,
we have
\begin{alignat*}{1}
\mathbb{E}\left[\left.Y_{i}\left(Z_{i}=1,\boldsymbol{x}_{i}\right)\right|\boldsymbol{x}_{i}\right] & =\mathbb{E}\left[\left.Y_{i}\left(Z_{i}=1,\boldsymbol{x}_{i}\right)\right|Z_{i}=1,\boldsymbol{x}_{i}^{\textrm{obs}}\right]+\zeta\left(Z_{i}=1,\boldsymbol{x}_{i}\right),\\
\mathbb{E}\left[\left.Y_{i}\left(Z_{i}=0,\boldsymbol{x}_{i}\right)\right|\boldsymbol{x}_{i}\right] & =\mathbb{E}\left[\left.Y_{i}\left(Z_{i}=0,\boldsymbol{x}_{i}\right)\right|Z_{i}=0,\boldsymbol{x}_{i}^{\textrm{obs}}\right]+\zeta\left(Z_{i}=0,\boldsymbol{x}_{i}\right),
\end{alignat*}
and 
\begin{alignat*}{1}
 & \mathbb{E}\left[\left.\zeta\left(Z_{i}=1,\boldsymbol{x}_{i}\right)\right|Z_{i}=1,\boldsymbol{x}_{i}^{\textrm{obs}}\right]=0,\\
 & \mathbb{E}\left[\left.\zeta\left(Z_{i}=0,\boldsymbol{x}_{i}\right)\right|Z_{i}=0,\boldsymbol{x}_{i}^{\textrm{obs}}\right]=0.
\end{alignat*}
Then, $\mathbb{E}\left[\left.Y_{i}\left(Z_{i}=1,\boldsymbol{x}_{i}\right)\right|Z_{i}=1,\boldsymbol{x}_{i}^{\textrm{obs}}\right]$
and $\mathbb{E}\left[\left.Y_{i}\left(Z_{i}=0,\boldsymbol{x}_{i}\right)\right|Z_{i}=0,\boldsymbol{x}_{i}^{\textrm{obs}}\right]$
are the $\boldsymbol{x}_{i}^{\textrm{obs}}$-projection model of the
observed outcome. The bias for each outcome is $\mathbb{E}\left[\left.\zeta\left(Z_{i}=1,\boldsymbol{x}_{i}\right)\right|\boldsymbol{x}_{i}^{\textrm{obs}}\right],$ $\mathbb{E}\left[\left.\zeta\left(Z_{i}=0,\boldsymbol{x}_{i}\right)\right|\boldsymbol{x}_{i}^{\textrm{obs}}\right]$
which satisfies
\begin{alignat*}{1}
 & \mathbb{E}\left[\left.\zeta\left(Z_{i}=1,\boldsymbol{x}_{i}\right)\right|\boldsymbol{x}_{i}^{\textrm{obs}}\right]\neq0,\\
 & \mathbb{E}\left[\left.\zeta\left(Z_{i}=0,\boldsymbol{x}_{i}\right)\right|\boldsymbol{x}_{i}^{\textrm{obs}}\right]\neq0.
\end{alignat*}
If we let the propensity score be, $\pi\left(\boldsymbol{x}_{i}\right)$,
this satisfies 
\[
\mathbb{E}\left[\left.\frac{\zeta\left(Z_{i}=1,\boldsymbol{x}_{i}\right)}{\pi\left(\boldsymbol{x}_{i}\right)}\right|\boldsymbol{x}_{i}^{\textrm{obs}}\right]=0,\ \mathbb{E}\left[\left.\frac{\zeta\left(Z_{i}=0,\boldsymbol{x}_{i}\right)}{1-\pi\left(\boldsymbol{x}_{i}\right)}\right|\boldsymbol{x}_{i}^{\textrm{obs}}\right]=0.
\]

While this bias is irreducible, we can define the ``best
approximate model" that achieves this bias using the (incomplete)
available information, $\boldsymbol{x}^{\textrm{obs}}$. We define the ``best approximate model" to be a model of the HTE, $\varphi\left(\boldsymbol{x}^{\textrm{obs}}\right)$, that satisfies
\begin{alignat}{1}\label{def:bam}
 & \mathbb{E}\left[\left.Y\left(Z=1,\boldsymbol{x}\right)\right|Z=1,\boldsymbol{x}^{\textrm{obs}}\right]-\mathbb{E}\left[\left.Y\left(Z=0,\boldsymbol{x}\right)\right|Z=0,\boldsymbol{x}^{\textrm{obs}}\right]\nonumber\\ 
= & \arg\min_{\varphi\in L^{2}\left(\boldsymbol{x}^{\textrm{obs}}\right)}\mathbb{E}\left[\left.\left(\tau\left(\boldsymbol{x}^{\textrm{obs}}\right)-\varphi\left(\boldsymbol{x}^{\textrm{obs}}\right)\right)^{2}\right|Z\right],
\end{alignat}
which has zero as its smallest value after removing the conditional $Z$.

It turns out that the best approximate model can also be expressed using a varying
coefficient model, as in Theorem~\ref{thm:VCM}.
First note that $\boldsymbol{x}^{\textrm{mis}}$ is unknown to all the estimators (i.e. no individual method has access to $\boldsymbol{x}^{\textrm{mis}}$).
In other words, all methods have access to $\boldsymbol{x}^{\textrm{obs}}$ but not $\boldsymbol{x}^{\textrm{mis}}$.
Since the availability of $\boldsymbol{x}^{\textrm{obs}}$ and unavailability of $\boldsymbol{x}^{\textrm{mis}}$ is the same for all $\boldsymbol{x}_{n+1}$,
if we take the conditional expectation, $\mathbb{E}\left[\left.X\right|\boldsymbol{x}_{n+1}^{\textrm{obs}}\right]$,
regarding $\boldsymbol{x}$, for both sides of the equation, we can
express the above as a function using $\boldsymbol{x}_{n+1}^{\textrm{obs}}$
only. From this, we have the following theorem.
\begin{thm}\label{thm:omit}
Let the estimator predictive values, $f_{j,n+1}\left(Z=1,\boldsymbol{x}_{n+1}^{\textrm{obs}}\right)$, $f_{j,n+1}\left(Z=0,\boldsymbol{x}_{n+1}^{\textrm{obs}}\right)$,
be a Gaussian field with $\boldsymbol{x}_{n+1}^{\textrm{obs}}$ as
parameters. 
Let $Y({1}),Y({0})$ be expressed as eq.~\eqref{eq:SVCM_2}, as with Theorem~\ref{thm:VCM}.
Since the $\boldsymbol{x}_{n+1}^{\textrm{obs}}$-conditional
expectation can be expressed with a measurable function with only $\boldsymbol{x}_{n+1}^{\textrm{obs}}$, if
we write each $\mathbb{E}_{x}\left[\left.\beta_{0}\left(\boldsymbol{x}_{n+1}\right)\right|\boldsymbol{x}_{n+1}^{\textrm{obs}}\right]$, $\left\{ \mathbb{E}_{x}\left[\left.\beta_{j}\left(\boldsymbol{x}_{n+1}\right)\right|\boldsymbol{x}_{n+1}^{\textrm{obs}}\right]\right\} _{j=1{:}J}$
as $\beta_{0}\left(\boldsymbol{x}_{n+1}^{\textrm{obs}}\right)$, $\left\{ \beta_{j}\left(\boldsymbol{x}_{n+1}^{\textrm{obs}}\right)\right\} _{1\leq j\leq J}$,
then we have 
\begin{equation}
\begin{alignedat}{1}\mathbb{E}\left[\left.Y\left(Z=1,\boldsymbol{x}_{n+1}\right)\right|\mathcal{G},\boldsymbol{x}_{n+1}^{\textrm{obs}},Z=1\right] & =\beta_{0}\left(Z=1,\boldsymbol{x}_{n+1}^{\textrm{obs}}\right)+\sum_{j=1}^{J}\beta_{j}\left(Z=1,\boldsymbol{x}_{n+1}^{\textrm{obs}}\right)f_{j,n+1}\left(Z=1,\boldsymbol{x}_{n+1}^{\textrm{obs}}\right),\\
\mathbb{E}\left[\left.Y\left(Z=0,\boldsymbol{x}_{n+1}\right)\right|\mathcal{G},\boldsymbol{x}_{n+1}^{\textrm{obs}},Z=0\right] & =\beta_{0}\left(Z=0,\boldsymbol{x}_{n+1}^{\textrm{obs}}\right)+\sum_{j=1}^{J}\beta_{j}\left(Z=0,\boldsymbol{x}_{n+1}^{\textrm{obs}}\right)f_{j,n+1}\left(Z=0,\boldsymbol{x}_{n+1}^{\textrm{obs}}\right),
\end{alignedat}
\label{eq:SVCM_3}
\end{equation}
where $\mathbb{E}\left[X\right]$ is the expectation operator
regarding the joint probability measure of $p_{x}$ and $\mu_{Y,\boldsymbol{f}}$. 
Thus, denoting the unobserved confounding bias as $\zeta\left(Z_{i}=1,\boldsymbol{x}_{i}\right),\zeta\left(Z_{i}=0,\boldsymbol{x}_{i}\right)$, we have
\begin{equation}
\begin{alignedat}{1}\mathbb{E}\left[\left.Y\left(Z=1,\boldsymbol{x}_{n+1}\right)\right|\mathcal{G},\boldsymbol{x}_{n+1}\right]-\mathbb{E}\left[\left.Y\left(Z=1,\boldsymbol{x}_{n+1}\right)\right|\mathcal{G},\boldsymbol{x}_{n+1}^{\textrm{obs}},Z=1\right] & =\zeta\left(Z_{i}=1,\boldsymbol{x}_{i}\right),\\
\mathbb{E}\left[\left.Y\left(Z=0,\boldsymbol{x}_{n+1}\right)\right|\mathcal{G},\boldsymbol{x}_{n+1}\right]-\mathbb{E}\left[\left.Y\left(Z=0,\boldsymbol{x}_{n+1}\right)\right|\mathcal{G},\boldsymbol{x}_{n+1}^{\textrm{obs}},Z=0\right] & =\zeta\left(Z_{i}=0,\boldsymbol{x}_{i}\right),
\end{alignedat}
\label{eq:SVCM_4}
\end{equation}
where the irreducible bias is achieved.
\end{thm}

This shows that, under unobserved confounding bias, our proposed method achieves the irreducible bias, and thus the best approximate model defined as eq.~\eqref{def:bam}.
While estimating the true HTE under unobserved confounding is impossible, eq.~(\ref{BPS}) is the closest we can get to the true HTE.

\subsection{Consistency of Bayesian causal synthesis}
In this section, we provide further theoretical justification of BCS.
Specifically, we derive its asymptotic behavior regarding its consistency in terms of estimating  the HTE.
Due to identification issues related to the synthesis model, the consistency of the HTE cannot be directly shown.
To circumvent this, we instead show predictive consistency of the treated and control outcomes, separately.
Then, given the predictive consistency of the two outcomes, we have consistency in the HTE as their difference.
While the results presented here are for when there are no unobserved confounding, if there exists unobserved confounding, we have consistency with regard to the true HTE plus irreducible bias.

Let $p\left(y_{n+1}\left|\boldsymbol{y}_{1{:}n},{x}_{1{:}n+1},Z_{1{:}n+1}\right.\right)$ denote the Bayesian predictive distribution generated from eq.~(\ref{BPS}), and $p\left(y_{n+1}\left|{x}_{n+1},Z_{n+1},\theta_{n+1}^{*}\right.\right)$ be the true distribution of the outcome.
Our goal is to construct a  predictive distribution, $p\left(y_{n+1}\left|\boldsymbol{y}_{1{:}n},{x}_{1{:}n+1},Z_{1{:}n+1}\right.\right)$, that is consistent with the target, $p\left(y_{n+1}\left|{x}_{n+1},Z_{n+1},\theta_{n+1}^{*}\right.\right)$.

Here, a consistent predictive distribution is defined as
\[
\lim_{n\rightarrow\infty}P\left(\left.\sup_{A\in\mathcal{B}\left(\mathbb{R}\right)}\left|p\left(y_{n+1}\in A\left|{x}_{n+1},Z_{n+1},\theta_{n+1}^{*}\right.\right)-p\left(y_{n+1}\in A\left|\boldsymbol{y}_{1{:}n},{x}_{1{:}n+1},Z_{1{:}n+1}\right.\right)\right|\right|Z\right)=0,
\]
where $P\left(\left.X\right|Z\right)$ is a probability measure of the DGP and $\mathcal{B}\left(\mathbb{R}\right)$ is a Borel set.  
We show a stronger result,
\[
\sup_{A\in\mathcal{B}\left(\mathbb{R}\right)}\left|p\left(y_{n+1}\in A\left|{x}_{n+1},Z_{n+1},\theta_{n+1}^{*}\right.\right)-p\left(y_{n+1}\in A\left|\boldsymbol{y}_{1{:}n},{x}_{1{:}n+1},Z_{1{:}n+1}\right.\right)\right|\rightarrow0,\ \textrm{in probability }P\left(\left.X\right|Z\right).
\]

\begin{thm}\label{thm:consistency}
Let $\mu^{*}\left({x},\pi\left({x}\right)\right)\in L^{2}\left(X\right)$,
$\boldsymbol{\beta}^{*}\left({x}\right)\in L^{2}\left(X\right)$
and the parameter space, $L^{2}\left(X\right)$, is complete and separable; i.e., a standard measure space.
Assume, $0<\pi(\mu^{*},\{ \boldsymbol{\beta}_{j}^{*}\}_{j=0,\cdots,J})$.
Then, we have
\begin{equation}
\lim_{n\rightarrow\infty}P\left(\left.\sup_{A\in\mathcal{B}\left(\mathbb{R}\right)}\left|p\left(y_{n+1}\in A\left|{x}_{n+1},Z_{n+1},\theta_{n+1}^{*}\right.\right)-p\left(y_{n+1}\in A\left|\boldsymbol{y}_{1{:}n},{x}_{1{:}n+1},Z_{n+1}\right.\right)\right|\right|Z\right)=0,
\label{eq:Consis}
\end{equation}
where $P\left(\left.X\right|Z\right)$ is a probability measure of eq.~\eqref{eq:VCM} and $\mathcal{B}\left(\mathbb{R}\right)$ is a Borel set. 
Similarly, we have
\begin{equation}
\sup_{A\in\mathcal{B}\left(\mathbb{R}\right)}\left|p\left(y_{n+1}\in A\left|{x}_{n+1},Z_{n+1},\theta_{n+1}^{*}\right.\right)-p\left(y_{n+1}\in A\left|\boldsymbol{y}_{1{:}n},{x}_{1{:}n+1},Z_{1{:}n+1}\right.\right)\right|\rightarrow0,\ \label{eq:TotalConsis}
\end{equation}
in probability $P\left(\left.X\right|Z\right)$.
\end{thm}

The proof of Theorem~\ref{thm:consistency} is in the Supplementary Material (Section~\ref{sec:proof-consistency}).
Theorem~\ref{thm:consistency} shows that the synthesis model (eq.~\ref{BPS}) is consistent, with regard to the predictive distribution of the outcome, given the treatment assignment, $Z$.
If there exists unobserved confounding, then the predictive difference will converge to the irreducible bias induced by unobserved confounding: $\zeta\left(Z_{i}=1,\boldsymbol{x}_{i}\right),\zeta\left(Z_{i}=0,\boldsymbol{x}_{i}\right)$.

The HTE estimate can be expressed by the expectation of the predictive distribution of $y_{n+1}$:
\[
{\tat}\left(X\right)=\int y_{n+1}\Big\{p\left(y_{n+1}\left|\boldsymbol{y}_{1{:}n},\boldsymbol{x}_{n+1},Z_{n+1}=1\right.\right)-p\left(y_{n+1}|\boldsymbol{y}_{1{:}n},\boldsymbol{x}_{1{:}n+1},Z_{1{:}n+1}=0\right)\Big\}dy_{n+1}.
\]
Since we have predictive consistency, if the convergence of the first moment is guaranteed, we can show that the synthesized estimate of the HTE is also consistent.

\begin{thm}
Assume that the predictive distributions, $p\left(y_{n+1}\in A\left|\boldsymbol{y}_{1{:}n},\boldsymbol{x}_{1{:}n+1},Z_{1{:}n+1}=1\right.\right)$ and $p\left(y_{n+1}\in A\left|\boldsymbol{y}_{1{:}n},\boldsymbol{x}_{1{:}n+1},Z_{1{:}n+1}=0\right.\right)$, are uniform integrable, that is,
\[
\lim_{c\rightarrow\infty}\limsup_{n\rightarrow\infty}\int_{\mathbb{R}}\left|y_{n+1}\right|1_{\left(\left|y_{n+1}\right|>c\right)}p\left(y_{n+1}\in A\left|\boldsymbol{y}_{1{:}n},\boldsymbol{x}_{1{:}n+1},Z_{1{:}n+1} \right.\right)dy_{n+1}=0
\]
for $Z_{n+1}\in\{0,1\}$.
Then, ${\tat}\left(X\right)\overset{\textrm{p}}{\rightarrow}\tau\left(X\right)$ holds. Therefore, the synthesized HTE estimate in eq.~(\ref{eq:tau}) is consistent.
\end{thm}
Again, if there exists unobserved confounding, we have the above result for ${\tat}\left(X\right)\overset{\textrm{p}}{\rightarrow}\tau\left(X\right)+\left(\zeta\left(Z=1,X\right)-\zeta\left(Z=0,X\right)\right)$, the true HTE plus the difference in irreducible bias of the outcomes.

\subsection{Further considerations}
There are two further reasons why this synthesis function is suitable for the task of estimating the HTE.
First, the BPS framework allows for flexible specification in the synthesis function.
In this case, the synthesis function has two crucial components for improved meta-inference.
The first component is the prognostic term, $\mu(X, \pi)$.
The inclusion of the propensity score $\pi\equiv \pi(X)$ in the prognostic term is known to improve the performance, as noted in \cite{hahn2020bayesian}, which we can include apart from the individual estimates.
Ensemble methods, in general, do not allow for this flexibility.

The second component is the averaging term, $\beta_0(X)+\sum_{j=1}^J \beta_j(X)f_j(X)$.
An important feature of eq.~\eqref{BPS} is that the weight, $\beta_j(X)$, of each HTE estimator can vary depending on the covariate, $X$.
This is crucial, as the performance of each estimator can vary depending on the subgroup/heterogeneity.
For example, one estimator might be good at estimating the treatment effect for the male subgroup, while another might be good for the female subgroup.
Typical ensemble methods (such as linear averaging or stacking), cannot take into account this heterogeneity in performance.
Comparatively, eq.~\eqref{BPS} can capture performance heterogeneity through $\beta_j(X)$.
Furthermore, $\beta_0(X)$ can capture additional variation in $\tau(X)$ that cannot be explained by the set of $J$ estimators, which aids in improving inference.

Second, because each estimate is treated as a latent factor in eq.~\eqref{eq:theorem1}, the posterior inference involves learning the biases and dependencies across estimates.
Because these estimates are similar, as they use the same data, similar procedures, or only differ in the selection of hyperparameters, the synthesis coefficients would have to take into account these characteristics to successfully infer.
Treating the HTE estimates as latent factors allow for this through Bayesian learning.

\section{Implementation and posterior computation \label{sec:comp}}

To complete the specifics of the above model (eq.~\ref{BPS}), we assume that the unknown functions $\mu(X, \pi)$ and $\beta_j(X)$ independently follow an isotropic Gaussian process. 
In particular, to assure computational scalability under large sample sizes, we employ a nearest-neighbor Gaussian process \citep{datta2016hierarchical}, that is, $\mu(X, \pi)\sim {\rm NNGP}_m(\barmu, \tau_{\mu}^2C(X; \phi_{\mu}))$ and $\beta_j(X)\sim {\rm NNGP}_m(\barbeta_j, \tau_{\beta_j}^2C(X;\phi_{\beta_j}))$, where $\tau_{\mu}^2$ and $\tau_{\beta_j}^2$ are unknown variance parameter, $\phi_{\mu}$ and $\phi_{\beta_j}$ are unknown spatial range parameter, $C(X;\phi)$ is a valid correlation function with spatial range parameter $\phi$, and $m$ is the number of nearest neighbors.
Here $\barbeta_j$ for $j=1,\ldots,J$ are global weight for the $j$-th method and also is a prior mean of the heterogeneous weight $\beta_j(X)$.
While it is possible to assign an informative prior on  $\barbeta_j$, we set $\barbeta_0=0$ and $\barbeta_j=1/J \ (j=1,\ldots,J)$, as a default choice, to make the prior synthesis function an equal weight averaging of the $J$ estimators. 

Given observed sample $(y_i, x_i, Z_i)$ for $i=1,\ldots,n$, the synthesis model (eq.~\ref{BPS}) for the observed sample is given by 
\begin{align*}
&y_i=\mu(x_i,\pi_i)+Z_i\left\{\beta_0(x_i)+\sum_{j=1}^J \beta_j(x_i)f_j(x_i)\right\}+\ep_i, \ \ \  \ep_i\sim N(0, \sigma^2),
\end{align*}
with $(\beta_j(x_1),\ldots, \beta_j(x_n))\sim N(\barbeta_j 1_n, \tau_{\beta_j}^2H_m(\phi_{\beta_j}; \mathcal{X}))$ independently for $j=0,1,\ldots,J$ and $(\mu(x_1,\pi_1),\ldots,\mu(x_n,\pi_n))\sim N(\barmu 1_n, \tau_\mu^2 H_m(\phi_\mu; \mathcal{Z}))$.
Here $H_m(X; \mathcal{X})$ and $H_m(X; \mathcal{Z})$ are $n\times n$ covariance matrices of the joint distribution on $n$ observations based on $m$-nearest neighbor Gaussian processes on $\mathcal{X}$ and $\mathcal{Z}$, respectively, where $\mathcal{X}$ and $\mathcal{Z}$ are space of $x_i$ and $z_i=(x_i,\pi_i)$, respectively. 
In what follows, we assign conditionally conjugate priors for unknown parameters other than spatial range parameters, that is, $\sigma^2\sim {\rm IG}(\delta_{\sigma}/2, \eta_{\sigma}/2)$, $\tau_\mu^2\sim {\rm IG}(\delta_{\mu}/2, \eta_{\mu}/2)$ and $\tau_{\beta_j}^2\sim {\rm IG}(\delta_{\beta_j}/2, \eta_{\beta_j}/2)$. 
For the spatial range parameter, we assign uniform priors, $\phi_\mu\sim U(\underline{c}_{\mu}, \overline{c}_{\mu})$ and $\phi_{\beta_j}\sim U(\underline{c}_{\beta}, \overline{c}_{\beta})$.
Then, the posterior distribution of the unknown parameters $\psi=(\sigma^2, \tau_\mu^2, \tau_{\beta_0}^2, \ldots,\tau_{\beta_J}^2, \phi_{\mu},\phi_{\beta_0},\ldots,\phi_{\beta_J})$, baseline function $\mu(X)$ and varying coefficients $\beta_0(X),\ldots,\beta_J(X)$ can be approximated by a Markov Chain Monte Carlo (MCMC) algorithm.  

To describe the MCMC algorithm, we let $\mu_i=\mu(z_i)$, $\beta_{ji}=\beta_j(x_i)$ and $f_{ji}=f_j(x_i)$, for notational simplicity. 
We define $\mu_i$ and $\Phi_n$ to be a collection of latent variables, namely, $\Phi_n=\{\mu_i, \beta_{0i},\ldots,\beta_{Ji}, f_{1i},\ldots,f_{Ji}\}_{i=1,\ldots,n}$.
The joint posterior distribution of $\psi$ and $\Phi_n$ is given by 
\begin{align*}
\Pi(\psi, \Phi_n|\mathcal{D})
&\propto \Pi(\psi)\prod_{i=1}^n \phi\Big(y_i; \mu_i+Z_i\big(\beta_{0i}+\sum_{j=1}^J \beta_{ji}f_{ji}\big),\sigma^2\Big)\prod_{j=1}^J h_j(f_{ji})\\
&  \ \ \ \ \ \ 
\times \phi_n(\mu^{(s)}; \barmu 1_n, \tau_\mu^2 H_m(\phi_\mu;\mZ))\prod_{j=0}^J \phi_n(\beta_j^{(s)}; \barbeta_j 1_n, \tau_{\beta_j}^2H_m(\phi_{\beta_j};\mX)),
\end{align*}
where $\mathcal{D}$ is a set of observed data, $\mu^{(s)}=(\mu_1,\ldots,\mu_n)$, $\beta_j^{(s)}=(\beta_{j1},\ldots,\beta_{jn})$ and $\Pi(\psi)$ is a joint prior distribution of $\psi$.
The posterior computation can be easily carried out via  Gibbs sampling, where step-by-step sampling steps are described in the Supplementary Material. 
Given the posterior samples of $\Phi_n$, the posterior samples of the synthesized HTE $\tau(x_i)$ given in eq.~\eqref{eq:tau} can be generated.

Regarding the inference on $\tau(x_0)$ for arbitrary $x_0\in \mX$, we suppose that $J$ estimators of $\tau(X_0)$ are available. 
From the distributional assumptions for $\beta_j(X)$, the conditional distribution of $\beta_j(x_0)$ is $\beta_j(x_0)|\beta_j^{(s)} \sim N(\barbeta_j + B_{\beta_j}(x_0)\beta_j(N(x_0)), \tau_{\beta_j}^2F_{\beta_j}(x_0) )$, where 
\begin{equation*}
\begin{split}
B_{\beta_j}(x_0)&=C_{\beta_j}(x_0, N(x_0); \phi_{\beta_j})C_{\beta_j}(N(x_0), N(x_0); \phi_{\beta_j})^{-1}, \\ 
F_{\beta_j}(x_0)&=1-C_{\beta_j}(x_0, N(x_0); \phi_{\beta_j})C_{\beta_j}(N(x_0), N(x_0); \phi_{\beta_j})^{-1}C_{\beta_j}(N(x_0), x_0; \phi_{\beta_j}),
\end{split}
\end{equation*}
and $N(x_0)$ denotes an index set of $m$-nearest neighbors of the evaluation point $x_0$.
Here $C_{\beta_j}(t, v; \phi_{\beta_j})$ for $t=(t_1,\ldots,t_{d_t})\in \mathbb{R}^{d_t}$ and $v=(v_1,\ldots,v_{d_v})\in \mathbb{R}^{d_v}$ denotes a $d_t\times d_v$ correlation matrix whose $(i_t, i_v)$-element is $C(\|Z_{i_t}-v_{i_v}\|; \phi_{\beta_j})$.
Then, given the posterior samples of $\beta_j$ and $\psi$, the posterior samples of $\beta_j(x_0)$ can be generated for $j=0,\ldots,J$.
This gives posterior samples of $\tau(x_0)$ provided that the HTE estimators at $x_0$, namely, $f_1(x_0),\ldots,f_J(x_0)$, are available.  

Detailed sampling procedure can be found in the Supplementary Material (Section~\ref{sec:NNGP-pos}).


\section{Simulation Study \label{sec:sim}}

To demonstrate and highlight the efficacy of our approach, we conduct a series of experiments.
The first simulation study in Section~\ref{sec:sim-syn} is based on synthetic data with unobserved confounders, and the second study in Section~\ref{sec:sim-acic} adopts several scenarios used in the Atlantic causal inference conference data analysis challenge.
Further simulation studies, including the simulation study in Section~\ref{sec:sim-syn} without unobserved confounders and a out-of-sample study, can be found in the Supplementary Material (Section~\ref{sec:supp-sim}).

\subsection{Synthetic data with unobserved confounders 
\label{sec:sim-syn}}

Let $X=(X_1,\ldots,X_p)$ be a $p$-dimensional (observed) covariate, where the fourth is a dichotomous variable, the fifth is an unordered categorical taking three levels (denoted 1, 2, 3), and the other variables are continuously drawn from standard normal distributions.
Further, let $W_1$ and $W_2$ be unobserved variables drawn from standard normal distributions.
The data generating process \citep[modified from][]{hahn2020bayesian} is 
$$
Y=\mu(X,W)+\tau(X,W)Z+\ep,  \ \ \ \ \ep\sim N(0, \sigma^2),
$$
where $T\sim {\rm Ber}(0.5W_1)$ is the treatment assignment and we set $\sigma^2=1$.
For the prognostic term $\mu(X,W)$ and treatment effect $\tau(X,W)$, we consider the following scenarios: 
\begin{align*}
&\text{(A)} \ \ \mu(X,W)=-7+6|X_3|-3X_5+\frac12W_1, \ \ \ \ \ \text{(B)} \ \ \mu(X,W)=2+2\sin(3X_3)+\frac12W_1\\
&\text{(A)} \ \ \tau(X,W)=1+2X_2X_5+W_2^2, \ \ \ \ \ \text{(B)} \ \ \tau(X,W)=1+2X_2X_5+\frac1 2 X_3^2+W_2^2,
\end{align*}
and we adopt four scenarios consisting of all the possible combinations. 
Note that we treat $W$ as ``randomized assignment" since we cannot observe $W_1$.

To estimate $\tau(X, W)$ without $W$, we consider the following methods:
\begin{itemize}
\item[-] 
{\bf Bayesian causal forest} \citep[BCF:][]{hahn2020bayesian}: Using R package (\url{https://github.com/socket778/XBCF}), apply the Bayesian causal forest to the observed data and generate 1500 posterior samples of each $\tau(X_i)$, after discarding the first 500 samples.

\item[-]
{\bf Linear model} (LM): Fitting a simple linear regression model, $Y=X^\top \beta_1+ZX^\top \beta_2+\ep$, and estimating $\tau(X)$ by $X^\top \hat{\beta}_2$.
Standard errors of the estimator are also computed. 

\item[-]
{\bf Additive models} (AM): Fitting an additive model, $Y=f_1(X)+Zf_2(X)+\ep$, and estimating $\tau(X)$ by $\hat{f}_2(X)$.
A crude approximation of the standard error of $\hat{f}_2(X)$ is computed by weighted bootstrap with 200 replications.

\item[-]
{\bf X-learner} \citep[XL:][]{kunzel2019metalearners}: Using R package (\url{https://github.com/xnie/rlearner}), apply X-learner with gradient boosting trees to the observed data to obtain point estimates of $\tau(X_i)$.

\item[-]
{\bf R-learner} \citep[RL:][]{nie2021quasi}: Using the same R package as X-learner, apply R-learner with gradient boosting trees to the observed data to obtain point estimates of $\tau(X_i)$.

\item[-]
{\bf Causal forest} \citep[CF:][]{wager2018estimation}: Using R package \texttt{grf}, apply the causal forest to the observed data and compute estimates and standard errors of $\tau(X_i)$ of the test data. 

\end{itemize}
Furthermore, we consider synthesizing three results (BCF, LM, AM) that have standard error estimates.
We adopt the following two synthesis methods: 
\begin{itemize}
\item[-]
{\bf Causal stacking} (CST): Applying the causal stacking algorithm by \cite{han2022ensemble} to ensemble the five methods given above, where the gradient boosting tree algorithm is used to construct prediction models in the averaging set. 
Note that this method only provides point estimates. 

\item[-]
{\bf Bayesian causal synthesis} (BCS): Apply the proposed method to synthesize the results of BCF, LM, and AM, and obtain posterior distributions of $\tau(X_i)$.
The number of nearest-neighbor is set to $m=15$. 
1500 posterior samples after discarding the first 500 samples are used to obtain posterior distributions of $\tau(X)$.
Note that only subsets of $X$ that are selected in AM are used for modeling the varying coefficients, $\beta_j(X)$. 
\end{itemize}

\begin{figure}[t!]
\centering
\includegraphics[width=13cm,clip]{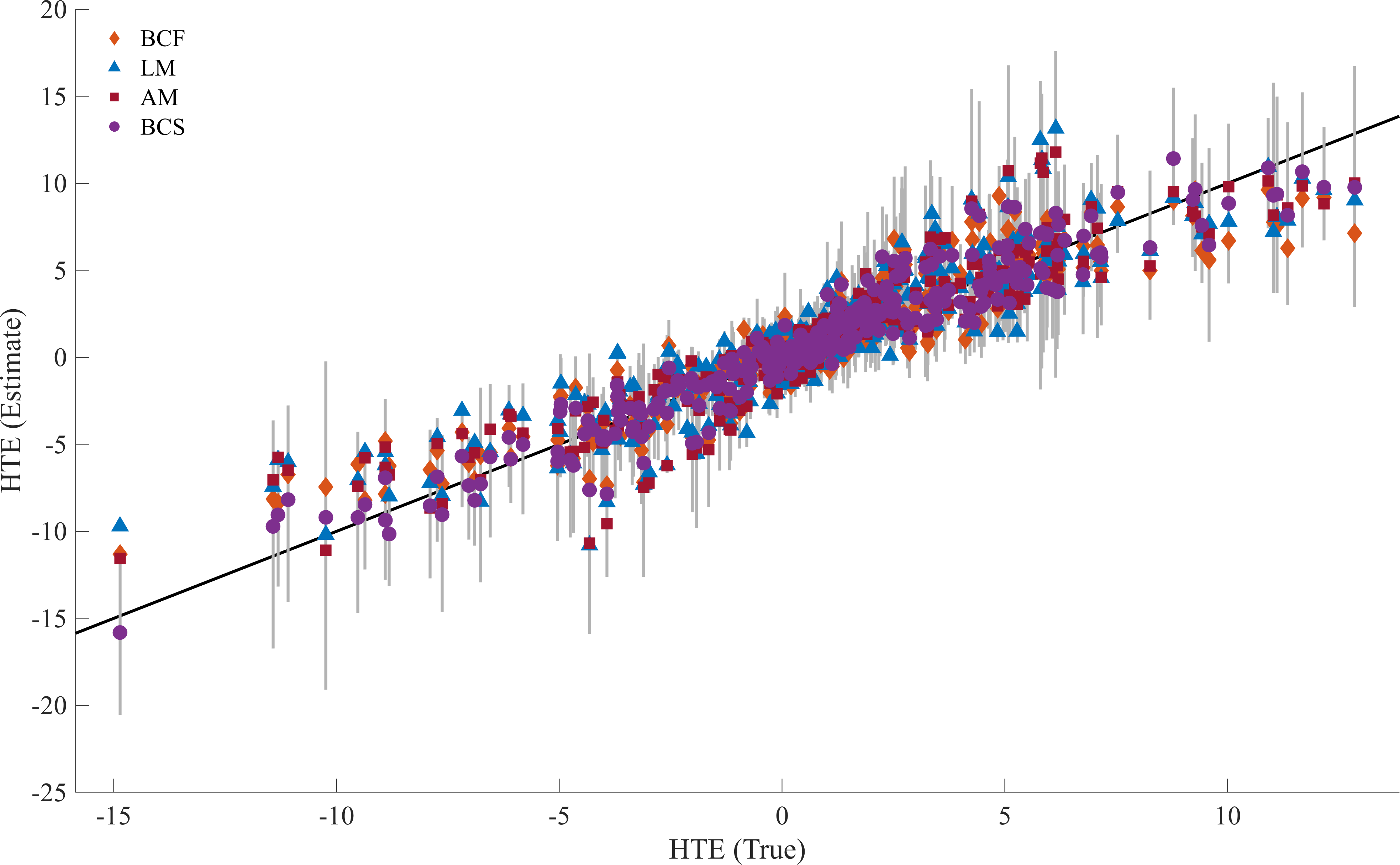}
\caption{Point estimates obtained from four methods and $95\%$ credible intervals of BCS (vertical line). 
\label{fig:oneshot1}
}
\end{figure}

\begin{figure}[t!]
\centering
\includegraphics[width=13cm,clip]{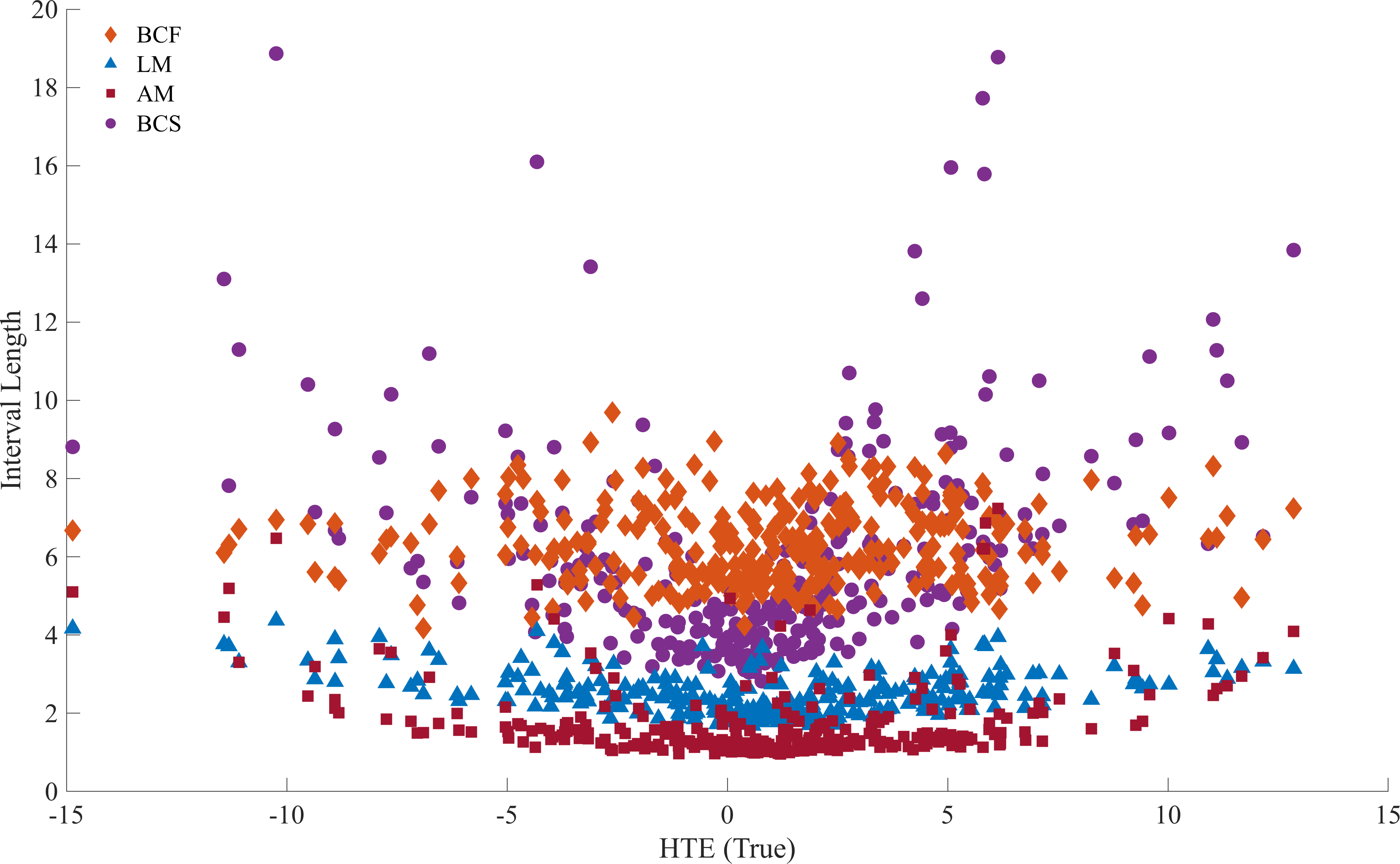}
\caption{Length of $95\%$ credible/confidence intervals against the true values of the heterogeneous treatment effect.
\label{fig:oneshot2}
}
\end{figure}

To demonstrate the effectiveness of BCS, we first show the results under the scenario  $(A)-(A)$.
The results of point prediction and $95\%$ credible intervals of BCS are presented in Figure~\ref{fig:oneshot1}, where the point predictions of the three models are also shown. 
BCS provides more accurate estimates than the three prediction methods, especially for samples having very small or large treatment effects. More importantly, the credible intervals of $\tau(X)$ obtained from BCS are $97.0\%$, which is only slightly larger than the nominal level of $95\%$.
To investigate the coverage property further, we present the interval lengths of credible (confidence) intervals of $\tau(X)$ in Figure~\ref{fig:oneshot2}. 
The result indicates that the posterior uncertainty on HTE around the region in which observed samples are abundant is small, while the uncertainty of the posterior obtained via BCS tends to be considerably large for regions where the observed samples are small. 
This is a reasonable and desirable result that reflects the uncertainty in each estimator, even though the other methods are overconfident.

\begin{figure}[t!]
\centering
\includegraphics[width=13cm,clip]{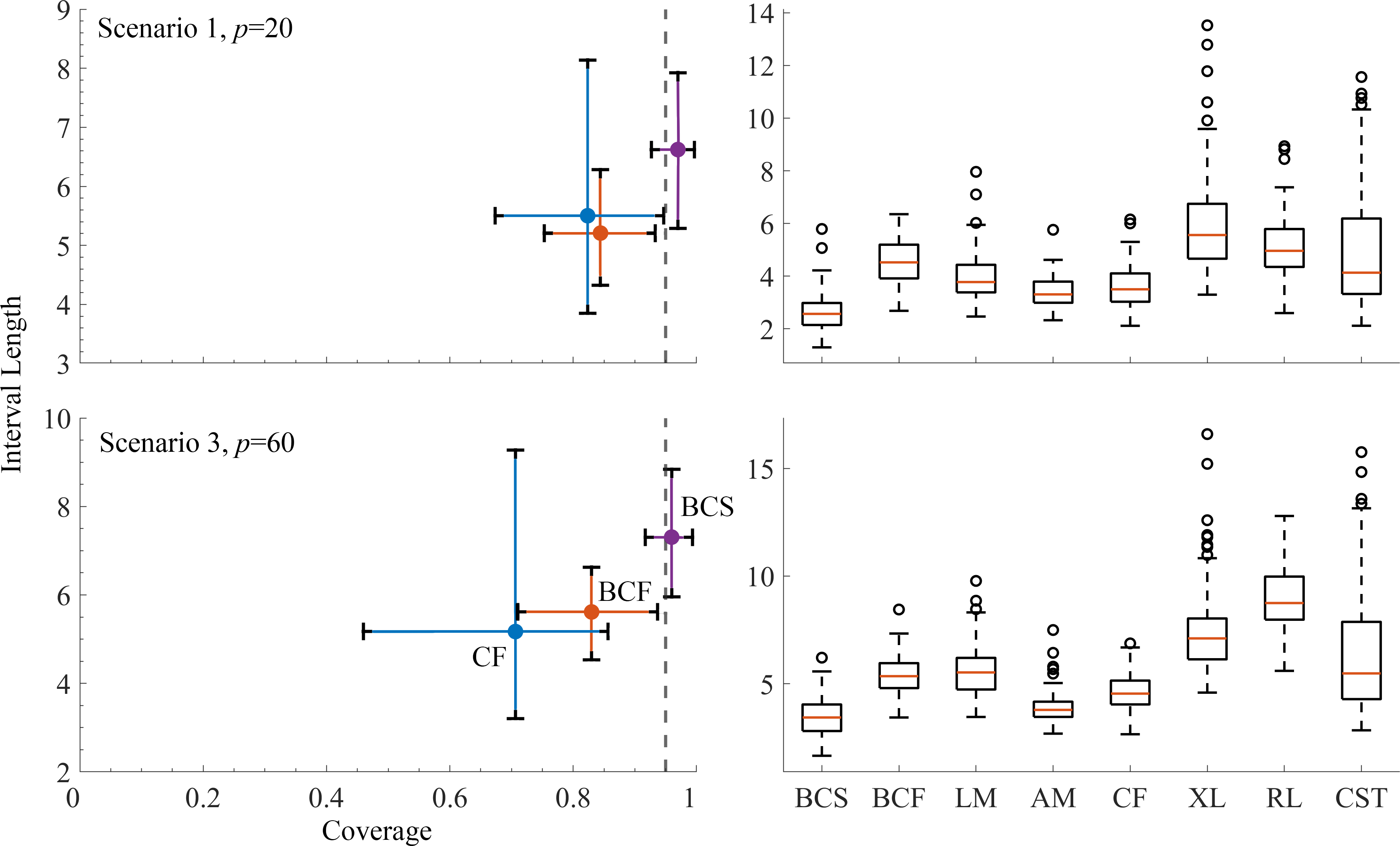}
\caption{Left column: Coverage probability (X-axis) and interval length (Y-axis) of the HTE estimates (top: Scenario 1, $p=20$; bottom: Scenario 3, $p=60$). The vertical and horizontal lines denote the 95\% interval. Right column: Boxplots of mean squared errors (MSE) of the estimation methods (top: Scenario 1, $p=20$; bottom: Scenario 3, $p=60$).
\label{fig:sim-MSE}
}
\end{figure}

We next evaluate the performance of the point and interval estimates through Monte Carlo replications. 
For the evaluation of point prediction, we compute the mean squared error (MSE):
$$
{\rm MSE}=\frac{1}{n}\sum_{i=1}^{n}\left\{\hat{\tau}(X_i)-\tau(X_i,Z_i)\right\}^2.
$$
In Figure~\ref{fig:sim-MSE}, we show the relationship between the CP and AL, and boxplots of MSE values of 100 Monte Carlo replications for two scenarios.
Overall, BCS provides more accurate estimates than the other methods, providing both improved point and interval estimates.
Looking at the interval estimates, we see that for Scenario 1, $p=20$,  BCS performs well, with the 95\% interval including the target coverage probability, even though BCF and CF  never achieves the nominal rate.
The conclusion is similar with Scenario 4, $p=60$, though the the performance gap has widened as the scenario has become more difficult.
For the point estimates, we see that BCS improves over all methods with significant gains in all scenarios.
Particular to note is that the median performance of BCS is better than the (non-outlier) minimum of many of the competing models.
Conversely, the non-outlier maximum of BCS is better than the median of several of the competing models.
This shows that BCS not only improves over the competing methods in terms of its median performance, but is robust to data variation.

Tables~\ref{tab:sim-smallmse} and \ref{tab:sim-smallcp} presents the MSE, and empirical coverage probability (CP) and average lengths (AL) of the $95\%$ credible (confidence) intervals of $\tau(X)$, obtained by BCS, BCF, LM, AM, XL, RL, CF,  and CST, while XL, RL, and CST do not provide uncertainty measures. 
The results show that 1) BCS outperforms the other methods for all scenarios, in terms of MSE, 2) generally, MSE increases as the dimension increases, 3) CST, given the aforementioned problem with standard ensemble methods, performs poorly compared to BCS, and 4) BCS produces the best CP compared to the nominal probability.
Overall, these results provide strong evidence for using BCS for estimating HTEs.

\begin{table}[t!]
\caption{Mean squared error (MSE) of point estimates of heterogeneous treatment effects, averaged over 100 Monte Carlo replications.
The smallest MSE values are highlighted in bold. 
}
\label{tab:sim-smallmse}
\begin{center}
\begin{tabular}{ccccccccccccc}
\hline
$\mu$ & $\tau$ & $p$ &  & BCS & BCF & LM & AM & CF & XL & RL & CST \\
\hline
 &  & 20 &  & {\bf 2.60} & 4.52 & 4.01 & 3.42 & 3.63 & 5.96 & 5.12 & 4.98 \\
(A) & (A) & 40 &  & {\bf 2.91} & 4.84 & 4.50 & 3.61 & 4.10 & 6.58 & 6.72 & 5.87 \\
 &  & 60 &  & {\bf 3.19} & 4.81 & 4.76 & 3.68 & 4.17 & 7.32 & 8.21 & 6.38 \\
 \hline
 &  & 20 &  & {\bf 2.95} & 4.34 & 3.40 & 3.51 & 3.36 & 5.22 & 4.55 & 4.61 \\
(B) & (A) & 40 &  & {\bf 3.02} & 4.50 & 3.58 & 3.46 & 3.62 & 5.74 & 5.69 & 5.94 \\
 &  & 60 &  & {\bf 3.07} & 4.54 & 3.63 & 3.59 & 3.77 & 6.48 & 6.80 & 6.73 \\
 \hline
 &  & 20 &  & {\bf 3.06} & 5.09 & 4.99 & 3.66 & 4.04 & 6.07 & 5.76 & 5.42 \\
(A) & (B) & 40 &  & {\bf 3.25} & 5.32 & 4.94 & 3.81 & 4.34 & 6.77 & 7.32 & 6.99 \\
 &  & 60 &  & {\bf 3.47} & 5.38 & 5.65 & 3.90 & 4.62 & 7.40 & 8.96 & 6.52 \\
 \hline
 &  & 20 &  & {\bf 3.01} & 4.42 & 3.75 & 3.59 & 3.51 & 5.47 & 4.91 & 5.20 \\
(B) & (B) & 40 &  & {\bf 3.32} & 4.70 & 4.14 & 3.74 & 3.96 & 6.29 & 6.08 & 6.15 \\
 &  & 60 &  & {\bf 3.41} & 4.85 & 4.20 & 3.73 & 4.17 & 6.80 & 7.35 & 6.37 \\
\hline
\end{tabular}
\end{center}
\end{table}

\begin{table}[t!]
\caption{Coverage probability (CP) and average length (AL) of $95\%$ credible/confidence intervals of heterogeneous treatment effects, averaged over 100 Monte Carlo replications.
}
\label{tab:sim-smallcp}
\begin{center}
\begin{tabular}{cccccccccccccccccc}
\hline
&&&& \multicolumn{4}{c}{CP ($\%$)} && \multicolumn{4}{c}{AL} \\
$\mu$ & $\tau$ & $p$ &  & BCS & BCF & LM & AM & CF &  & BCS & BCF & LM & AM & CF \\
\hline
 &  & 20 &  & 97.0 & 84.4 & 54.3 & 51.2 & 82.3 &  & 6.62 & 5.21 & 2.81 & 2.45 & 5.50 \\
(A) & (A) & 40 &  & 97.0 & 84.0 & 53.9 & 52.1 & 75.0 &  & 6.94 & 5.37 & 3.03 & 2.57 & 4.77 \\
 &  & 60 &  & 96.6 & 84.0 & 52.8 & 53.4 & 70.5 &  & 7.06 & 5.49 & 3.07 & 2.66 & 4.75 \\
 \hline
 &  & 20 &  & 93.4 & 83.8 & 45.3 & 47.3 & 80.5 &  & 5.88 & 4.92 & 1.97 & 2.28 & 4.78 \\
(B) & (A) & 40 &  & 92.9 & 83.9 & 45.2 & 48.1 & 71.0 &  & 5.83 & 5.05 & 2.08 & 2.31 & 3.79 \\
 &  & 60 &  & 93.3 & 84.0 & 43.5 & 48.3 & 67.0 &  & 5.97 & 5.15 & 2.06 & 2.43 & 3.70 \\
 \hline
 &  & 20 &  & 96.5 & 82.9 & 55.4 & 55.3 & 80.7 &  & 7.06 & 5.44 & 3.20 & 2.75 & 5.56 \\
(A) & (B) & 40 &  & 96.4 & 82.7 & 55.0 & 54.9 & 73.4 &  & 7.20 & 5.53 & 3.22 & 2.80 & 5.03 \\
 &  & 60 &  & 96.0 & 83.0 & 56.6 & 54.1 & 70.6 &  & 7.31 & 5.61 & 3.61 & 2.76 & 5.17 \\
 \hline
 &  & 20 &  & 92.6 & 82.8 & 44.8 & 51.8 & 77.1 &  & 5.95 & 4.96 & 2.09 & 2.53 & 4.59 \\
(B) & (B) & 40 &  & 91.1 & 82.0 & 44.9 & 51.0 & 68.8 &  & 5.98 & 5.10 & 2.28 & 2.55 & 4.02 \\
 &  & 60 &  & 91.1 & 82.7 & 44.3 & 49.9 & 63.3 &  & 6.00 & 5.26 & 2.27 & 2.49 & 3.68 \\
\hline
\end{tabular}
\end{center}
\end{table}

\subsection{Atlantic causal inference conference data analysis challenge 2017 \label{sec:sim-acic}}
Atlantic causal inference conference (ACIC) data analysis challenge is a prominent causal inference competition, where several methods are compared in a blind manner.
In the competition, there are four data generating processes, but we focus on the most complicated underlying structures (called ``non-additive errors"), where the detailed description of the data generating process is given in \cite{hahn2019atlantic}. 
We consider 8 scenarios, as presented in \cite{hahn2019atlantic}. 
We randomly sampled $n=500$ samples from the simulated data, and applied the same methods in Section~\ref{sec:sim-syn}.
Based on $R=100$ replications, the performance is evaluated through root mean squared errors (RMSE), given by 
$$
{\rm RMSE}=\left[\frac{1}{nR}\sum_{r=1}^R\sum_{i=1}^n\left\{\hat{\tau}^{(r)}(X_i)-\tau^{(r)}(X_i)\right\}^2\right]^{1/2},
$$
where $\hat{\tau}^{(r)}(X_i)$ and $\tau^{(r)}(X_i)$ are the estimated and true values of the heterogeneous treatment effect, respectively, in the $r$-th replication. 
The results are reported in Table \ref{tab:sim-ACIC1}.
For BCS, BCF, and CF, we also assess the empirical coverage probability (CP) and average length (AL) of $95\%$ credible or confidence intervals. 
The results are given in Table \ref{tab:sim-ACIC2}.

Looking at the point estimates (Table \ref{tab:sim-ACIC1}), we see that BCS outperforms every other method. Given that BCF (and their variants) are top performers in the competition, improving upon BCF with a clear margin is notable.
Focusing on the CP (Table \ref{tab:sim-ACIC2}), we see the true strengths of BCF.
Specifically, even though the other methods, including BCF, produce subpar CPs, deviating from the nominal 95\%, BCS is able to produce CPs that are within approximately 3\% of the nominal 95\%.
Considering the difficulty of the scenarios, achieving such an accurate CP is a good indication of the performance of BCS.

\begin{table}[t!]
\caption{Root mean squared errors of point estimates of heterogeneous treatment effects of observed samples with the ACIC data, averaged over 100 replications. 
The smallest MSE values are highlighted in bold. 
}
\label{tab:sim-ACIC1}
\begin{center}
\begin{tabular}{ccccccccccccc}
\hline
Scenario & & BCS & BCF & LM & AM & CF & XL & RL & CST \\
\hline
1 &  & {\bf 1.04} & 2.31 & 1.87 & 1.89 & 1.34 & 1.08 & 1.30 & 1.64 \\
2 &  & {\bf 1.10} & 2.44 & 1.88 & 1.90 & 1.77 & 1.31 & 1.69 & 1.61 \\
3 &  & {\bf 1.31} & 2.18 & 1.86 & 1.88 & 1.58 & 1.40 & 1.62 & 1.59 \\
4 &  & {\bf 1.45} & 2.41 & 1.89 & 1.94 & 2.02 & 1.63 & 2.13 & 1.77 \\
5 &  & {\bf 1.00} & 2.32 & 1.85 & 1.87 & 1.32 & 1.08 & 1.29 & 1.68 \\
6 &  & {\bf 1.10} & 2.50 & 1.87 & 1.90 & 1.81 & 1.34 & 1.73 & 1.59 \\
7 &  & {\bf 1.41} & 1.97 & 1.86 & 1.88 & 1.77 & 1.52 & 1.78 & 1.69 \\
8 &  & {\bf 1.54} & 2.14 & 1.93 & 1.98 & 2.14 & 1.78 & 2.30 & 1.83 \\
\hline
\end{tabular}
\end{center}
\end{table}

\begin{table}[t!]
\caption{Coverage probability (CP) and average length (AL) of $95\%$ credible (confidence) intervals of heterogeneous treatment effects of observed samples with the ACIC data, averaged over 100 replications.  
}
\label{tab:sim-ACIC2}
\begin{center}
\begin{tabular}{ccccccccccc}
\hline
&  & \multicolumn{1}{c}{}    & \multicolumn{1}{c}{CP}  & \multicolumn{1}{c}{}   & \multicolumn{1}{c}{} & \multicolumn{1}{c}{}    & \multicolumn{1}{c}{AL}  & \multicolumn{1}{c}{}   \\
Scenario  &  & BCS & BCF & CF &  & BCS & BCF & CF \\
\hline
1 &  & 97.2 & 67.2 & 81.1 &  & 4.14 & 4.21 & 2.75 \\
2 &  & 95.9 & 55.5 & 79.6 &  & 4.10 & 3.61 & 4.06 \\
3 &  & 97.7 & 78.2 & 88.9 &  & 5.86 & 5.06 & 5.13 \\
4 &  & 96.3 & 73.2 & 85.5 &  & 6.02 & 5.31 & 6.40 \\
5 &  & 97.5 & 63.5 & 83.0 &  & 4.34 & 4.05 & 2.89 \\
6 &  & 96.4 & 56.4 & 78.8 &  & 4.51 & 3.97 & 4.33 \\
7 &  & 97.1 & 84.8 & 87.6 &  & 6.22 & 5.49 & 5.65 \\
8 &  & 96.5 & 76.2 & 83.9 &  & 6.45 & 5.50 & 6.57 \\
\hline
\end{tabular}
\end{center}
\end{table}

\section{Application: The Effect of Smoking on Medical Expenditures \label{sec:app}}

\begin{figure}[t!]
\centering
\includegraphics[width=13cm,clip]{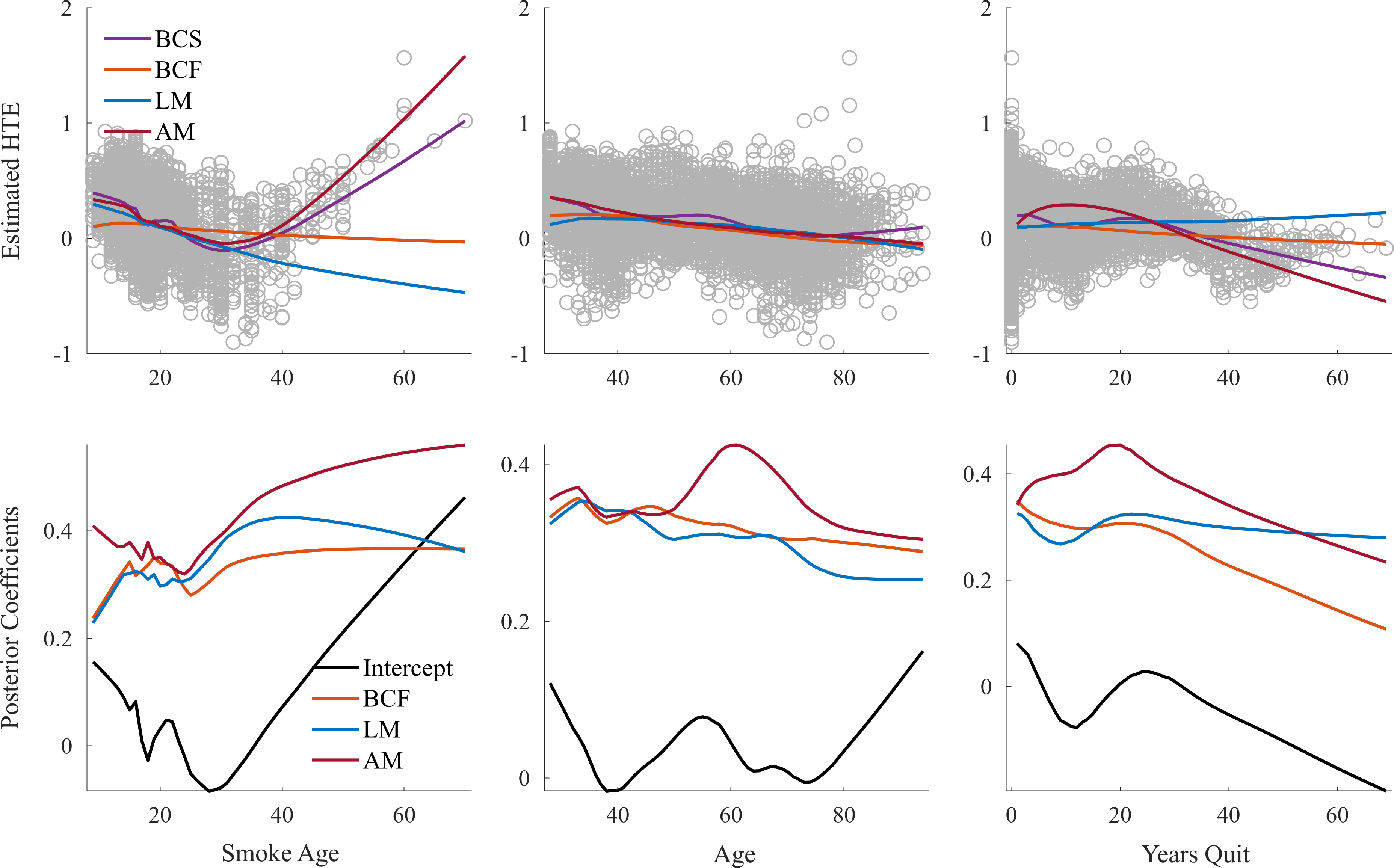}
\caption{The estimated treatment effects (top) and model coefficients (bottom) against three continuous covariates. 
\label{fig:app1}
}
\end{figure}

As an empirical demonstration, we consider the question of how smoking affects medical expenditures. 
This question has been studied in several previous papers; though here,
we follow \cite{imai2004causal} in analyzing data extracted from the 1987 National Medical Expenditure Survey (NMES) by Johnson et al. (2003). 
Our setup includes the following ten patient attributes:
\begin{itemize}
\itemsep-1em 
    \item[-] \texttt{age}: age in years at the time of the survey
    \item[-] \texttt{smoke age}: age in years when the individual started smoking
    \item[-] \texttt{gender}: male or female
    \item[-] \texttt{race}: other, black or white
    \item[-] \texttt{marriage status}: married, widowed, divorced, separated, never married
    \item[-] \texttt{education level}: college graduate, some college, high school graduate, other
    \item[-] \texttt{census region}: geographic location, Northeast, Midwest, South, West
    \item[-] \texttt{poverty status}: poor, near poor, low income, middle income, high income
    \item[-] \texttt{seat belt}: does the patient regularly use a seat belt when in a car
    \item[-] \texttt{years quit}: how many years since the individual quit smoking
\end{itemize}
We estimate the treatment effect for \texttt{smoke age}, \texttt{age}, and \texttt{years quit}. The results are given in Figure~\ref{fig:app1}.
For BCS, we synthesize BCF, LM, and AM.

Looking at the top row in Figure~\ref{fig:app1}, we find several interesting features.
For one, the HTE profile is quite different between the three underlying methods, and across covariates.
Specifically, LM can be seen to be too rigid, having the linearity assumption, compared to BCF and LAM, which are more flexible.
Focusing specifically on \texttt{Smoke Age}, we see that the heterogeneity between the three underlying methods to be quite different.
BCS, taking the best of all worlds, can be seen to balance the three methods well.

Moving the attention to the BCS coefficients (Figure~\ref{fig:app1}, bottom row), some interesting patterns highlight the strengths of BCS.
For example, looking at \texttt{Smoke Age}, we see a significant increase in the intercept as smoke age increases.
As mentioned previously, the intercept captures the misspecification that the totality of the estimators cannot capture (i.e. the model set misspecification).
For smoke age, there are a lot fewer people who start smoking later in life, and thus fewer people in those subgroups.
This causes considerable misspecification and uncertainty.
The increase in the intercept as the age increases reflects this.
This pattern is also seen in the other subgroups, but to a lesser extent due to those subgroups having more members.
Looking at the coefficients of each method, interestingly, AM has the most weight for most cases.
This is surprising, since we would expect BCF to perform the best, and thus have the most weight.
We also observe some regions where there is a reversal in coefficients, most notably the latter half of \texttt{Years Quit}, where there is a significant drop in coefficients for AM and BCF, while LM remains constant.
This shows that BCS is able to capture performance heterogeneity amongst subgroups.

While an in-depth analysis of the application is beyond the scope of this paper, it is clear that the inference obtained from BCS is sufficiently different from the other methods.
This has broad implications for other studies, as BCS can provide a more robust and accurate estimate of the HTE.

\section{Summary Comments \label{sec:summ}}
Working under a coherent Bayesian framework that provides a theoretically and conceptually sound way to synthesize information, we develop Bayesian causal synthesis, a method to synthesize multiple causal estimates.
With this new method, decision makers can calibrate, learn, and update coefficients on each estimation method, and improve estimation.
We show, through theoretical analysis, that BCS provides HTE estimates that are consistent.
Several simulation studies, including those used in causal inference competitions, show that BCS, by synthesizing several HTE estimates, outperforms other estimation methods, in terms of mean squared error and coverage probability.
A real data observational study also highlights the flexibility, interpretability, and efficacy of BCS.

In addition to the causal problems addressed in this paper, there are several areas in which BCS can be useful.
This includes RCTs and other experiments, as well as observational studies where temporal or spatial information is relevant.
Another clear extension is to meta-analysis, where experimental information can be used to synthesize estimates effectively.
We believe that the flexibility of our approach will lead to further developments that will define increasing empirical support for the utility and efficacy of the approach, and attract applied researchers.

\section*{Acknowledgement}
This work is partially supported by Japan Society for Promotion of Science (KAKENHI) grant numbers 21H00699.

\bibliographystyle{chicago}
\bibliography{refs}

\newpage
\setcounter{equation}{0}
\setcounter{section}{0}
\setcounter{table}{0}
\setcounter{page}{1}
\renewcommand{\thesection}{S\arabic{section}}
\renewcommand{\theequation}{S\arabic{equation}}
\renewcommand{\thetable}{S\arabic{table}}

\vspace{1cm}
\begin{center}
{\LARGE
{\bf Supplementary Material for ``Bayesian Causal Synthesis for Meta-Inference on Heterogeneous Treatment Effect"}
}
\end{center}

This Supplementary Material provides the proofs of the theorems presented in the paper, the posterior sampling algorithm,  and additional simulation results.

\section{Proofs}

\subsection{Proof of Theorem ~\ref{thm:VCM}\label{supp:th1}}

We provide the proof for $Y\left(Z=1,X\right)$. The proof follows
identically for $Y\left(Z=0,X\right)$. 

\subsubsection{Preliminary}

Let $H_{x}^{\otimes\left(J+1\right)}$ denote the Hilbert space constructed
from the $J+1$ vector valued functions, where each component of the
vector is an element of $H_{x}$. The inner product regarding $\left(H_{x}^{\otimes\left(J+1\right)}\ni\right)\boldsymbol{g}=\left(g^{1},\cdots,g^{J+1}\right)^{\top}$,
$H_{x}^{\otimes\left(J+1\right)}$ is defined as 
\[
\left\langle \boldsymbol{g},\boldsymbol{h}\right\rangle _{H_{x}^{\otimes\left(J+1\right)}}=\mathbb{E}_{x}\left[\sum_{j=1}^{J+1}g^{j}\left(\boldsymbol{x}\right)h^{j}\left(\boldsymbol{x}\right)\right].
\]

Let $\mu_{Y({1}),\boldsymbol{f}}$ be the joint probability measure
of $\left(Y\left(Z=1,X\right),\left\{ f_{j}\left(Z=1,X\right)\right\} _{j=1{:}J}\right)$.
Here, $\mu_{Y({1}),\boldsymbol{f}}$ is a probability measure on $H_{x}^{\otimes\left(J+1\right)}$,
where the marginals, 
\[
\left(Y\left(Z=1,X\right),\left\{ f_{j}\left(Z=1,X\right)\right\} _{j=1{:}J}\right)
\]
are each $\mu_{Y({1})},\left\{ \mu_{f_{j}({1})}\right\} _{j=1{:}J}$.
Let $m_{Y({1}),\boldsymbol{f}}=\left[m_{Y({1})},m_{f_{1}},\cdots,m_{f_{J}}\right]^{\top}$
be the mean vector and $R_{Y({1}),\boldsymbol{f}}$ the covariance
operator of the Gaussian measure, $\mu_{Y({1}),\boldsymbol{f}}$,
on $H_{x}^{\otimes\left(J+1\right)}$. Here, $\mu_{Y({1}),\boldsymbol{f}},\mu_{Y({1})},\left\{ \mu_{f_{j}}\right\} _{j=1{:}J}$
are all Gaussian measures, and, for example, the tensor, $\mu_{Y({1})}\otimes\mu_{f_{j}}$,
of the marginals, $\mu_{Y({1})},\left\{ \mu_{f_{j}}\right\} _{j=1{:}J}$,
are also Gaussian measures. If we set the positive definite kernel,
$\kappa_{R}:H_{x}^{\otimes\left(J+1\right)}\times H_{x}^{\otimes\left(J+1\right)}\rightarrow\mathbb{R}$,
as 
\[
\kappa_{R}\left(\boldsymbol{g},\boldsymbol{h}\right)=\left\langle R_{Y,\boldsymbol{f}}\boldsymbol{g},\boldsymbol{h}\right\rangle _{H_{x}^{\otimes\left(J+1\right)}},\ \boldsymbol{g},\boldsymbol{h}\in H^{\otimes\left(J+1\right)},
\]
then we can construct the RKHS, $\mathcal{H}_{\kappa}^{\otimes\left(J+1\right)}$,
that corresponds to $\kappa_{R}$. Then, there exists a CONS (complete
orthonormal system), $\left\{ \boldsymbol{e}_{k};k\in\mathbb{N}\right\} $,
of $\mathcal{H}_{\kappa}^{\otimes\left(J+1\right)}$, where $e_{k}$
is a $\mathbb{R}^{J+1}$-valued function, $\left(\boldsymbol{e}_{k}=\left[e_{k}^{Y},e_{k}^{1},\cdots,e_{k}^{J}\right]^{\top}\right)$,
such that $R_{Y,\boldsymbol{f}}\boldsymbol{e}_{k}=\lambda_{k}\boldsymbol{e}_{k}$
has corresponding eigenvalues, $\lambda_{k}$, and eigenfunctions,
$\boldsymbol{e}_{k}$. The Gaussian random variable, $Z\left(X\right)$,
that takes values on $H_{x}^{\otimes\left(J+1\right)}$, can be expanded
as 
\begin{equation}
Z\left(X\right)=m_{Y({1}),\boldsymbol{f}}\left(X\right)+\sum_{k=1}^{\infty}\sqrt{\lambda_{k}}\frac{\left\langle Z-m_{Y({1}),\boldsymbol{f}},\boldsymbol{e}_{k}\right\rangle _{\mathcal{H}_{\kappa}^{\otimes\left(J+1\right)}}}{\sqrt{\lambda_{k}}}\boldsymbol{e}_{k}\left(X\right),\ \textrm{a.s. }\mu,\label{eq:GaussianSeries}
\end{equation}
where $\left\langle X,X\right\rangle _{\mathcal{H}_{\kappa}^{\otimes\left(J+1\right)}}$
is the inner product regarding $\mathcal{H}_{\kappa}^{\otimes\left(J+1\right)}$,
and $\left\{ \frac{1}{\sqrt{\lambda_{k}}}\left\langle Z,\boldsymbol{e}_{k}\right\rangle _{\mathcal{H}_{\kappa}^{\otimes\left(J+1\right)}}:k\in\mathbb{N}\right\} $
are independent, for each $k$, standard Gaussian random variable
sequence: $\frac{\left\langle Z,\boldsymbol{e}_{k}\right\rangle }{\sqrt{\lambda_{k}}}\sim\mathcal{N}\left(0,1\right)$.
This means that the information regarding the joint probability measure,
$\mu_{Y\left(1\right),\boldsymbol{f}}$, is fully aggregated in $\left(m_{Y({1}),\boldsymbol{f}},R_{Y,\boldsymbol{f}}\right)$
and $\left(m_{Y({1}),\boldsymbol{f}},\lambda_{k},\boldsymbol{e}_{k}\right)$.
Let $\boldsymbol{e}_{k}^{j}$ denote $\boldsymbol{e}_{k}^{j}=\left[\cdots,0,e_{k}^{j},0\cdots\right]^{\top}$.
Then, when $i\neq j$, we have $\left\langle e_{k}^{i},e_{k}^{j}\right\rangle _{\mathcal{H}_{\kappa}^{\otimes\left(J+1\right)}}=0$.

\subsubsection{Proof of Theorem ~\ref{thm:VCM}}

If we series expand the DGP, $Y\left(Z=1,X\right)$, as 
\[
Y\left(Z=1,X\right)=m_{Y({1})}\left(X\right)+\sum_{k=1}^{\infty}\left\langle Z-m_{Y({1}),\boldsymbol{f}},\boldsymbol{e}_{k}^{Y}\right\rangle _{\mathcal{H}_{\kappa}^{\otimes\left(J+1\right)}}\boldsymbol{e}_{k}^{Y}\left(X\right),
\]
this is an element of the Gaussian space (a space of independent countable
infinite Gaussian vectors), $\textrm{span}\left\{ \left\langle Z-m_{Y({1}),\boldsymbol{f}},\boldsymbol{e}_{k}^{Y}\right\rangle _{\mathcal{H}_{\kappa}^{\otimes\left(J+1\right)}};k\in\mathbb{N}\right\} $.
The estimator predictive values, $\left\{ f_{j,n+1}\left(Z=1,X\right)\right\} _{j=1{:}J}$,
is 
\[
f_{j,n+1}\left(Z=1,X\right)=m_{j}\left(X\right)+\sum_{k=1}^{\infty}\left\langle Z-m_{Y({1}),\boldsymbol{f}},\boldsymbol{e}_{k}^{j}\right\rangle _{\mathcal{H}_{\kappa}^{\otimes\left(J+1\right)}}e_{k}^{j}\left(X\right),
\]
and the generated Gaussian space, $\mathfrak{F}$, from the estimator
predictive values, $\left\{ f_{j}\left(Z=1,X\right)\right\} _{j=1{:}J}$,
is $\mathfrak{F}=\bigcup_{j=1}^{J}\textrm{span}\left\{ \left\langle Z-m_{Y({1}),\boldsymbol{f}},\boldsymbol{e}_{k}^{j}\right\rangle _{\mathcal{H}_{\kappa}^{\otimes\left(J+1\right)}};k\in\mathbb{N}\right\} $.
From the Gaussianity of $\mu_{Y\left(1\right),\boldsymbol{f}}$, $\mathbb{E}_{\mu_{Y\left(1\right),\boldsymbol{f}}}\left[Y\left|\mathcal{F}\right.\right]$
is equivalent to a $H$-valued Gaussian random variable, $Y\left(1\right)$,
projecting to the Gaussian space, $\mathfrak{F}$. Therefore, if we
denote $P_{\mathfrak{F}}$ as the orthogonal projection operator of
$Y\left(1\right)$ to $\mathfrak{F}$, we have 
\[
P_{\mathfrak{F}}\left(Z-m_{Y({1}),\boldsymbol{f}}\right)=\mathbb{E}_{\mu_{Y,f}}\left[Z-m_{Y({1}),\boldsymbol{f}}\left|\mathcal{F}\right.\right].
\]

We now show that the best approximate model for $Y\left(Z=1,X\right)$,
given the realizations from the estimator probability field of each point,
$\left\{ f_{j}\left(\boldsymbol{x}_{l}\right);1\leq l\leq n+1\right\} _{1\leq j\leq J}$,
is a covariate varying coefficient model.

Since the value, $\left\{ f_{j}\left(\boldsymbol{x}_{l}\right);1\leq l\leq n+1\right\} $
can be considered as 
\[
\left\{ \sum_{k=1}^{\infty}\left\langle Z,\boldsymbol{e}_{k}^{j}\right\rangle _{\mathcal{H}_{\kappa}^{\otimes\left(J+1\right)}}e_{k}^{j}\left(\boldsymbol{x}_{l}\right);1\leq l\leq n+1\right\} ,
\]
$\mathcal{G}$ generated from the realization, $\left\{ f_{j}\left(\boldsymbol{x}_{l}\right);1\leq l\leq n+1\right\} _{1\leq j\leq J}$,
is 
\[
\mathcal{G}=\mathscr{B}\left(\textrm{span}\left\{ \sum_{k=1}^{\infty}\left\langle Z,\boldsymbol{e}_{k}^{j}\right\rangle _{\mathcal{H}_{\kappa}^{\otimes\left(J+1\right)}}e_{k}^{j}\left(\boldsymbol{x}_{l}\right);1\leq l\leq n+1\right\} _{j=1{:}J}\right).
\]
If we let $\mathfrak{G}$ be the Gaussian space corresponding to $\mathcal{G}$,
we have 
\[
\mathfrak{G}=\bigcup_{j=1}^{J}\textrm{span}\left\{ \sum_{k=1}^{\infty}e_{k}^{j}\left(\boldsymbol{x}_{l}\right)\left\langle Z-m_{Y({1}),\boldsymbol{f}},\boldsymbol{e}_{k}^{j}\right\rangle _{\mathcal{H}_{\kappa}^{\otimes\left(J+1\right)}};1\leq l\leq n+1\right\} ,
\]
where $\mathfrak{G}\subset\mathfrak{F}$. The conditional expectation,
$\mathbb{E}_{\mu_{Y,f}}\left[\left.Y\left(Z=1,X\right)\right|\mathcal{G}\right]$,
can be expressed as \citep[][Theorem 3.10.1]{Bogachev_98},
\begin{alignat}{1}
 & \mathbb{E}_{\mu_{Y,f}}\left[\left.Y\left(Z=1,X\right)-m_{Y({1})}\left(X\right)\right|\mathcal{G}\right]\label{eq:Conditional-1}\\
= & P_{\mathfrak{G}}\left(\sum_{i=1}^{\infty}\left\langle Z-m_{Y({1}),\boldsymbol{f}},\boldsymbol{e}_{i}^{Y}\right\rangle _{\mathcal{H}_{\kappa}^{\otimes\left(J+1\right)}}e_{i}^{Y}\left(X\right)\right)\nonumber \\
= & \sum_{i=1}^{\infty}\sum_{j=1}^{J}\sum_{l=1}^{n+1}c_{il}^{j}\left(\sum_{k=1}^{\infty}e_{k}^{j}\left(\boldsymbol{x}_{l}\right)\left\langle Z-m_{Y({1}),\boldsymbol{f}},\boldsymbol{e}_{k}^{j}\right\rangle _{\mathcal{H}_{\kappa}^{\otimes\left(J+1\right)}}\right)e_{i}^{Y}\left(X\right)\nonumber 
\end{alignat}
using a double subscript sequence, $\left\{ c_{il}^{j}\right\} _{i,l\in\mathbb{N}}^{1\leq j\leq J}$,
such that $\sum_{i=1}^{\infty}\sum_{l=1}^{\infty}\left|c_{il}^{j}\right|^{2}<\infty$.

Since, $f_{j}\left(\boldsymbol{x}_{l}\right)-m_{f_{j}}\left(\boldsymbol{x}_{l}\right)=\sum_{k=1}^{\infty}e_{k}^{j}\left(\boldsymbol{x}_{l}\right)\left\langle Z-m_{Y({1}),\boldsymbol{f}},\boldsymbol{e}_{k}^{j}\right\rangle _{\mathcal{H}_{\kappa}^{\otimes\left(J+1\right)}}$,
the value of $Y$ we want to predict at $\boldsymbol{x}_{n+1}$, $\mathbb{E}\left[\left.Y\left(\boldsymbol{x}_{n+1}\right)\right|\mathcal{G}\right]$,
is 
\begin{alignat*}{1}
 & m_{Y({1})}\left(\boldsymbol{x}_{n+1}\right)+\sum_{i=1}^{\infty}\sum_{j=1}^{J}\sum_{l=1}^{n+1}c_{il}^{j}\left(f_{j}\left(\boldsymbol{x}_{l}\right)-m_{f_{j}}\left(\boldsymbol{x}_{l}\right)\right)e_{i}^{Y}\left(\boldsymbol{x}_{n+1}\right).
\end{alignat*}
If we rearrange the terms 
\begin{alignat*}{1}
m_{Y({1})}\left(\boldsymbol{x}_{n+1}\right)+\sum_{j=1}^{J}\sum_{l=1}^{n}\sum_{i=1}^{\infty}c_{il}^{j}\left(f_{j}\left(\boldsymbol{x}_{l}\right)-m_{f_{j}}\left(\boldsymbol{x}_{l}\right)\right)e_{i}^{Y}\left(\boldsymbol{x}_{n+1}\right)\\
+\sum_{j=1}^{J}\sum_{i=1}^{\infty}c_{i,n+1}^{j}\left(f_{j}\left(\boldsymbol{x}_{n+1}\right)-m_{f_{j}}\left(\boldsymbol{x}_{n+1}\right)\right)e_{i}^{Y}\left(\boldsymbol{x}_{n+1}\right),
\end{alignat*}
and denote $\beta_{0}\left(\boldsymbol{x}_{n+1}\right),\beta_{j}\left(\boldsymbol{x}_{n+1}\right)$
as 
\begin{alignat*}{1}
\beta_{0}\left(\boldsymbol{x}_{n+1}\right) & =m_{Y({1})}\left(\boldsymbol{x}_{n+1}\right)+\sum_{j=1}^{J}\sum_{l=1}^{n}\sum_{i=1}^{\infty}c_{il}^{j}\left(f_{j}\left(\boldsymbol{x}_{l}\right)-m_{f_{j}}\left(\boldsymbol{x}_{l}\right)\right)e_{i}^{Y}\left(\boldsymbol{x}_{n+1}\right)\\
 & -\sum_{j=1}^{J}\sum_{i=1}^{\infty}c_{i,n+1}^{j}m_{f_{j}}\left(\boldsymbol{x}_{n+1}\right)e_{i}^{Y}\left(\boldsymbol{x}_{n+1}\right),\\
\beta_{j}\left(\boldsymbol{x}_{n+1}\right) & =\sum_{i=1}^{\infty}c_{i,n+1}^{j}e_{i}^{Y}\left(\boldsymbol{x}_{n+1}\right),
\end{alignat*}
from (\ref{eq:Conditional-1}), the random variable, $\mathbb{E}_{\mu_{Y\left(1\right),\boldsymbol{f}}}\left[\left.Y\left(\boldsymbol{x}_{n+1}\right)\right|\mathcal{G}\right]$,
at $\boldsymbol{x}_{n+1}$ is 
\begin{alignat*}{1}
\mathbb{E}_{\mu_{Y\left(1\right),\boldsymbol{f}}}\left[\left.Y\left(\boldsymbol{x}_{n+1}\right)\right|\mathcal{G}\right] & =\beta_{0}\left(\boldsymbol{x}_{n+1}\right)+\sum_{j=1}^{J}\beta_{j}\left(\boldsymbol{x}_{n+1}\right)f_{j}\left(\boldsymbol{x}_{n+1}\right),\ \left(^{\forall}s\in S\right).
\end{alignat*}
The probability field, $Y\left(Z=1,X\right)$, is 
\[
Y\left(Z=1,X\right)=\mathbb{E}_{\mu_{Y\left(1\right),\boldsymbol{f}}}\left[\left.Y\left(Z=1,X\right)\right|\mathcal{G}\right]+\varepsilon\left(X\right),\ \left(\varepsilon\left(X\right)\textrm{ is independent of }\mathcal{G}\right),
\]
which means that this can be expressed using a covariate varying coefficient
model, 
\begin{alignat*}{1}
Y\left(\boldsymbol{x}_{n+1}\right) & =\beta_{0}\left(\boldsymbol{x}_{n+1}\right)+\sum_{j=1}^{J}\beta_{j}\left(\boldsymbol{x}_{n+1}\right)f_{j}\left(\boldsymbol{x}_{n+1}\right)+\varepsilon\left(\boldsymbol{x}_{n+1}\right),\ \mathbb{E}_{\mu_{Y,f}}\left[\left.\varepsilon\left(\boldsymbol{x}_{n+1}\right)\right|\mathcal{G}\right]=0,\ \left(^{\forall}\boldsymbol{x}_{n+1}\right).
\end{alignat*}

\subsection{Proof of Theorem \ref{thm:omit}}

Since, $f_{j}\left(\boldsymbol{x}_{l}^{\textrm{obs}}\right)-m_{f_{j}}\left(\boldsymbol{x}_{l}^{\textrm{obs}}\right)=\sum_{k=1}^{\infty}e_{k}^{j}\left(\boldsymbol{x}_{l}^{\textrm{obs}}\right)\left\langle Z-m_{Y({1}),\boldsymbol{f}},\boldsymbol{e}_{k}^{j}\right\rangle _{\mathcal{H}_{\kappa}^{\otimes\left(J+1\right)}}$,
the value of $Y$ we want to predict at $\boldsymbol{x}_{n+1}\left(=\left[\begin{array}{c}
\boldsymbol{x}_{n+1}^{\textrm{obs}}\\
\mathbf{x}_{n+1}^{\textrm{unobs}}
\end{array}\right]\right)$, $\mathbb{E}\left[\left.Y\left(\boldsymbol{x}_{n+1}\right)\right|\mathcal{G}\right]$,
is 
\begin{alignat*}{1}
 & m_{Y({1})}\left(\boldsymbol{x}_{n+1}\right)+\sum_{i=1}^{\infty}\sum_{j=1}^{J}\sum_{l=1}^{n+1}c_{il}^{j}\left(f_{j}\left(\boldsymbol{x}_{l}^{\textrm{obs}}\right)-m_{f_{j}}\left(\boldsymbol{x}_{l}^{\textrm{obs}}\right)\right)e_{i}^{Y}\left(\boldsymbol{x}_{n+1}\right).
\end{alignat*}
If we rearrange the terms 
\begin{alignat*}{1}
m_{Y({1})}\left(\boldsymbol{x}_{n+1}\right)+\sum_{j=1}^{J}\sum_{l=1}^{n}\sum_{i=1}^{\infty}c_{il}^{j}\left(f_{j}\left(\boldsymbol{x}_{l}^{\textrm{obs}}\right)-m_{f_{j}}\left(\boldsymbol{x}_{l}^{\textrm{obs}}\right)\right)e_{i}^{Y}\left(\boldsymbol{x}_{n+1}\right)\\
+\sum_{j=1}^{J}\sum_{i=1}^{\infty}c_{i,n+1}^{j}\left(f_{j}\left(\boldsymbol{x}_{n+1}^{\textrm{obs}}\right)-m_{f_{j}}\left(\boldsymbol{x}_{n+1}^{\textrm{obs}}\right)\right)e_{i}^{Y}\left(\boldsymbol{x}_{n+1}\right),
\end{alignat*}
and denote $\beta_{0}\left(\boldsymbol{x}_{n+1}\right),\beta_{j}\left(\boldsymbol{x}_{n+1}\right)$
as 
\begin{alignat*}{1}
\beta_{0}\left(\boldsymbol{x}_{n+1}\right) & =m_{Y({1})}\left(\boldsymbol{x}_{n+1}\right)+\sum_{j=1}^{J}\sum_{l=1}^{n}\sum_{i=1}^{\infty}c_{il}^{j}\left(f_{j}\left(\boldsymbol{x}_{l}^{\textrm{obs}}\right)-m_{f_{j}}\left(\boldsymbol{x}_{l}^{\textrm{obs}}\right)\right)e_{i}^{Y}\left(\boldsymbol{x}_{n+1}\right)\\
 & -\sum_{j=1}^{J}\sum_{i=1}^{\infty}c_{i,n+1}^{j}m_{f_{j}}\left(\boldsymbol{x}_{n+1}^{\textrm{obs}}\right)e_{i}^{Y}\left(\boldsymbol{x}_{n+1}\right),\\
\beta_{j}\left(\boldsymbol{x}_{n+1}\right) & =\sum_{i=1}^{\infty}c_{i,n+1}^{j}e_{i}^{Y}\left(\boldsymbol{x}_{n+1}\right),
\end{alignat*}
from (\ref{eq:Conditional-1}), the random variable, $\mathbb{E}_{\mu_{Y\left(1\right),\boldsymbol{f}}}\left[\left.Y\left(\boldsymbol{x}_{n+1}\right)\right|\mathcal{G}\right]$,
at $\boldsymbol{x}_{n+1}$ is 
\begin{alignat*}{1}
\mathbb{E}_{\mu_{Y\left(1\right),\boldsymbol{f}}}\left[\left.Y\left(\boldsymbol{x}_{n+1}\right)\right|\mathcal{G}\right] & =\beta_{0}\left(\boldsymbol{x}_{n+1}\right)+\sum_{j=1}^{J}\beta_{j}\left(\boldsymbol{x}_{n+1}\right)f_{j}\left(\boldsymbol{x}_{n+1}^{\textrm{obs}}\right),\ \left(^{\forall}s\in S\right).
\end{alignat*}
The probability field, $Y\left(Z=1,X\right)$, is 
\[
Y\left(Z=1,X\right)=\mathbb{E}_{\mu_{Y\left(1\right),\boldsymbol{f}}}\left[\left.Y\left(Z=1,X\right)\right|\mathcal{G}\right]+\varepsilon\left(X\right),\ \left(\varepsilon\left(X\right)\textrm{ is independent of }\mathcal{G}\right),
\]
which means that this can be expressed using a covariate varying coefficient
model, 
\begin{alignat*}{1}
Y\left(\boldsymbol{x}_{n+1}\right) & =\beta_{0}\left(\boldsymbol{x}_{n+1}\right)+\sum_{j=1}^{J}\beta_{j}\left(\boldsymbol{x}_{n+1}\right)f_{j}\left(\boldsymbol{x}_{n+1}^{\textrm{obs}}\right)+\varepsilon\left(\boldsymbol{x}_{n+1}\right),\ \mathbb{E}_{\mu_{Y,f}}\left[\left.\varepsilon\left(\boldsymbol{x}_{n+1}\right)\right|\mathcal{G}\right]=0,\ \left(^{\forall}\boldsymbol{x}_{n+1}\right).
\end{alignat*}

Since, this expression holds for all $\boldsymbol{x}_{n+1}$, if we
take the conditional expectation, $\mathbb{E}_{x}\left[\left.X\right|\boldsymbol{x}_{n+1}^{\textrm{obs}}\right]$,
regarding $\boldsymbol{x}$, for both sides, we have 
\begin{alignat*}{1}
\mathbb{E}_{x}\left[\left.Y\left(\boldsymbol{x}_{n+1}\right)\right|\boldsymbol{x}_{n+1}^{\textrm{obs}}\right] & =\mathbb{E}_{x}\left[\left.\beta_{0}\left(\boldsymbol{x}_{n+1}\right)\right|\boldsymbol{x}_{n+1}^{\textrm{obs}}\right]+\sum_{j=1}^{J}\mathbb{E}_{x}\left[\left.\beta_{j}\left(\boldsymbol{x}_{n+1}\right)\right|\boldsymbol{x}_{n+1}^{\textrm{obs}}\right]f_{j}\left(\boldsymbol{x}_{n+1}^{\textrm{obs}}\right)\\
 & +\mathbb{E}_{x}\left[\left.\varepsilon\left(\boldsymbol{x}_{n+1}\right)\right|\boldsymbol{x}_{n+1}^{\textrm{obs}}\right].
\end{alignat*}
Here, $\mathbb{E}_{x}\left[\left.\beta_{0}\left(\boldsymbol{x}_{n+1}\right)\right|\boldsymbol{x}_{n+1}^{\textrm{obs}}\right],\left\{ \mathbb{E}_{x}\left[\left.\beta_{j}\left(\boldsymbol{x}_{n+1}\right)\right|\boldsymbol{x}_{n+1}^{\textrm{obs}}\right]\right\} _{1\leq j\leq J}$
is a function that can be solely expressed using $\boldsymbol{x}_{n+1}^{\textrm{obs}}$,
and $\mathbb{E}_{x}\left[\left.\varepsilon\left(\boldsymbol{x}_{n+1}\right)\right|\boldsymbol{x}_{n+1}^{\textrm{obs}}\right]$
is 
\[
\mathbb{E}_{\mu_{Y\left(1\right),\boldsymbol{f}}}\left[\left.\mathbb{E}_{x}\left[\left.\varepsilon\left(\boldsymbol{x}_{n+1}\right)\right|\boldsymbol{x}_{n+1}^{\textrm{obs}}\right]\right|\mathcal{G}\right]=0.
\]
Therefore, if we write $\mathbb{E}_{x}\left[\left.\beta_{0}\left(\boldsymbol{x}_{n+1}\right)\right|\boldsymbol{x}_{n+1}^{\textrm{obs}}\right],\left\{ \mathbb{E}_{x}\left[\left.\beta_{j}\left(\boldsymbol{x}_{n+1}\right)\right|\boldsymbol{x}_{n+1}^{\textrm{obs}}\right]\right\} _{1\leq j\leq J}$
with each $\beta_{0}\left(\boldsymbol{x}_{n+1}^{\textrm{obs}}\right)$,
$\left\{ \beta_{j}\left(\boldsymbol{x}_{n+1}^{\textrm{obs}}\right)\right\} _{1\leq j\leq J}$,
we have 
\begin{alignat*}{1}
\mathbb{E}\left[\left.Y\left(Z=1,\boldsymbol{x}_{n+1}\right)\right|\mathcal{F}_{t},\boldsymbol{x}_{n+1}^{\textrm{obs}}\right] & =\beta_{0}\left(Z=1,\boldsymbol{x}_{n+1}^{\textrm{obs}}\right)+\sum_{j=1}^{J}\beta_{j}\left(Z=1,\boldsymbol{x}_{n+1}^{\textrm{obs}}\right)f_{j,n+1}\left(Z=1,\boldsymbol{x}_{n+1}^{\textrm{obs}},0\right),\\
\mathbb{E}\left[\left.Y\left(Z=0,\boldsymbol{x}_{n+1}\right)\right|\mathcal{F}_{t},\boldsymbol{x}_{n+1}^{\textrm{obs}}\right] & =\beta_{0}\left(Z=0,\boldsymbol{x}_{n+1}^{\textrm{obs}}\right)+\sum_{j=1}^{J}\beta_{j}\left(Z=0,\boldsymbol{x}_{n+1}^{\textrm{obs}}\right)f_{j,n+1}\left(Z=0,\boldsymbol{x}_{n+1}^{\textrm{obs}},0\right).
\end{alignat*}

\subsection{Proof of Theorem~\ref{thm:consistency}}

\label{sec:proof-consistency} 

\subsubsection{Preliminary}

We assume that the following varying coefficient model, 
\begin{alignat}{1}
y({x}_{i}) & =\mu^{*}\left({x}_{i},\pi\right)+Z\left({x}_{i}\right)\left\{ \beta_{0}^{*}\left({x}_{i}\right)+\sum_{j=1}^{J}\beta_{j}^{*}\left({x}_{i}\right)f_{j}\left({x}_{i}\right)\right\} +\varepsilon_{i},\ \ \ \ i=1,\ldots,n,\label{eq:VCM}
\end{alignat}
with $\mathbb{E}\left[\varepsilon_{i}\right]=0$ and $\mathbb{E}\left[\varepsilon_{i}^{2}\right]=\sigma_{i}^{2}$,
is correctly specified as a regression function. Thus, we assume that
$\mathbb{E}\left[\left.\varepsilon_{i}\right|x_{i}\right]=0$ holds
for $i=1,\ldots,n$. Further, for the propensity score, $\pi$, we
assume $0\ll\pi\left(X\right)\ll1$, for all $X$. This means that
for all treatment hierarchies, there exist subjects for assignment/non-assignment.

Denote $y_{i}=y\left({x}_{i}\right)$, $\mu_{i}=\mu\left({x}_{i},\pi\right)$,
$Z_{i}=Z\left({x}_{i}\right)$, $f_{ji}=f_{j}\left({x}_{i}\right)$,
$\beta_{ji}^{*}=\beta_{j}^{*}\left({x}_{i}\right)$ as the actual
value of the random variable concerning the explanatory variable,
${x}_{i}$. $\boldsymbol{\mu}^{n}=\left(\mu_{1},\cdots,\mu_{n}\right)^{\top}$,
$\boldsymbol{\beta}_{j}^{n}=\left(\beta_{j1},\cdots,\beta_{jn}\right)^{\top}$,
$\boldsymbol{f}_{j}^{n}=\left(f_{j1},\cdots,f_{jn}\right)^{\top}$.
With regard to the parameters, we denote the $n$-th parameter vector
as $\boldsymbol{\theta}^{n}=\left(\mu^{n},\boldsymbol{\beta}^{n},\sigma^{2}\right)$.
Here, $\sigma^{2}I_{n\times n}$ is the covariance matrix of $\left\{ \varepsilon_{i}\right\} _{i=1,\cdots n}$.
The observation to be predicted, $y_{n+1}\left({x}_{n+1}\right)$,
given the explanatory variable, ${x}_{n+1}$, can be written as 
\begin{alignat}{1}
y_{n+1} & =\mu_{n+1}^{*}+Z_{n+1}\left\{ \beta_{0,n+1}^{*}+\sum_{j=1}^{J}\beta_{j,n+1}^{*}f_{j,n+1}\right\} +\varepsilon_{n+1},\ \ \varepsilon_{n+1}\sim N\left(0,\sigma^{2}\right)\label{eq:VCMahead}\\
 & \boldsymbol{\beta}_{n+1}^{*}=\left(\beta_{1,n+1}^{*},\cdots,\beta_{J,n+1}^{*}\right)^{\top},\ \ \left\Vert \boldsymbol{\beta}_{n+1}^{*}\right\Vert _{L^{2}\left(X\right)}<\infty.\nonumber 
\end{alignat}
The unknown parameter set is $\theta_{n+1}^{*}=\left(\mu_{n+1}^{*},\boldsymbol{\beta}_{n+1}^{*},\sigma^{2}\right)$,
and the distribution of eq.~\eqref{eq:VCMahead} is written as $p\left(y_{n+1}\left|{x}_{n+1},Z_{n+1},\theta_{n+1}^{*}\right.\right)$.

We now construct the predictive distribution of eq.~\eqref{eq:VCMahead},
using a Gaussian process. Let the prior distributions be $\mu\sim\textrm{GP}\left(\tau_{\mu},h_{\mu}\right)$,
$\boldsymbol{\beta}_{j}\sim\textrm{GP}\left(\tau_{j},h_{j}\right)$
and $\pi\left(\sigma\right)=\sigma^{-1}$. Here, we assume that $\boldsymbol{\beta}_{j}$
are independent for $j$. The conditional likelihood function of $\boldsymbol{y}_{1{:}n}$,
given $\{\boldsymbol{f}_{j}^{n}\}_{j=1\cdots J}$ and $\mathcal{L}(\boldsymbol{y}_{1{:}n}|\boldsymbol{\theta}^{n},T,\{\boldsymbol{f}_{j}^{n}\}_{j=1\cdots J})$,
is $N\big(\boldsymbol{y}_{1{:}n}-\boldsymbol{\mu}^{n}-T(\boldsymbol{\beta}_{0}^{n}+\sum_{j=1}^{J}\boldsymbol{\beta}_{j}^{n}\circ\boldsymbol{f}_{j}^{n}),\sigma^{2}I_{n\times n})$,
where the $\circ$ operator is the Hadamard product. The likelihood
function $\mathcal{L}\left(\boldsymbol{y}_{1{:}n}\left|\boldsymbol{\theta}^{n},T\right.\right)$,
is convoluted by the estimator density, $\left\{ h_{j}\right\} $, which
gives us 
\begin{alignat*}{1}
\mathcal{L}\left(\boldsymbol{y}_{1{:}n}\left|\boldsymbol{\theta}^{n},T\right.\right) & =\int_{\mathbb{R}^{J}}N\left(\boldsymbol{y}_{1{:}n}-\boldsymbol{\mu}^{n}-T\Big(\boldsymbol{\beta}_{0}^{n}+\sum_{j=1}^{J}\boldsymbol{\beta}_{j}^{n}\circ\boldsymbol{f}_{j}^{n}\Big),\sigma^{2}I_{n\times n}\right)\prod_{j=1}^{J}h_{j}\left(\left.f_{j}^{n}\right|x^{n}\right)df.
\end{alignat*}
Then, the conditional joint probability density of $\boldsymbol{y}^{n+1}=((\boldsymbol{y}_{1{:}n})^{\top},y_{n+1})^{\top}$
given ${x}_{1{:}n+1}$ is 
\begin{alignat}{1}
P\left(\left.\boldsymbol{y}^{n+1}\right|{x}_{1{:}n+1},T\right)= & \int_{\mathbb{R}^{J+2}}\mathcal{L}\left(\boldsymbol{y}^{n+1}\left|\boldsymbol{\theta}^{n+1},T\right.\right)\pi\left(\boldsymbol{\theta}^{n+1}\right)d\boldsymbol{\theta}^{n+1}\label{eq:Predictive}
\end{alignat}
and the predictive distribution of $y_{n+1}$ is 
\[
p\left(y_{n+1}\left|\boldsymbol{y}_{1{:}n},{x}_{1{:}n+1},T\right.\right)=\frac{P\left(\left.\boldsymbol{y}^{n+1}\right|{x}_{1{:}n+1},T\right)}{\int_{\mathbb{R}}P\left(\left.\boldsymbol{y}^{n+1}\right|{x}_{1{:}n+1},T\right)dy_{n+1}}.
\]
In what follows, we show that the above predictive distribution is
consistent with the true distribution $p\left(y_{n+1}\left|{x}_{n+1},Z_{n+1},\theta_{n+1}^{*}\right.\right)$.

\subsubsection{Proof of Theorem \ref{thm:consistency}}
\begin{proof}
This proof is for the $Z=1$ case, though the proof for $Z=0$ follows
equivalently. Let the direct product measure (cylinder measure) of
the data, $\left(y_{1},\cdots,y_{n},\cdots\right)$, be $\mathbb{P}_{\infty}^{*}$,
and the direct product measure (cylinder measure) of the data of the
synthesis model, $\left(y_{1},\cdots,y_{n},\cdots\right)$, be $\mathbb{P}_{\infty}=\mathcal{L}\left(\boldsymbol{y}^{\infty}\left|\boldsymbol{\theta}^{\infty}\right.\right)$.
Denote the marginal under the Gaussian process prior as 
\[
\mathbb{P}_{\infty}\pi^{\otimes\infty}=\int_{\mathbb{R}^{J\otimes\infty},\mathbb{R}_{+}}\mathcal{L}\left(\boldsymbol{y}^{\infty}\left|\boldsymbol{\theta}^{\infty}\right.\right)\pi\left(\boldsymbol{\theta}^{\infty}\right)d\boldsymbol{\theta}^{\infty}.
\]
As with $\mathbb{P}_{n}^{*},\mathbb{P}_{n}\pi^{\otimes n}$, denote
the conditional distribution of the cylinder measure, given $\boldsymbol{y}_{1{:}n}$,
as $\left.\mathbb{P}_{\infty}^{*}\right|_{\boldsymbol{y}_{1{:}n}},\left.\mathbb{P}_{\infty}\pi^{\otimes\infty}\right|_{\boldsymbol{y}_{1{:}n}}$.
Note that 
\[
\left.\mathbb{P}_{\infty}^{*}\right|_{\boldsymbol{y}_{1{:}n}}=\frac{\mathcal{L}\left(\boldsymbol{y}^{\infty}\left|\boldsymbol{\theta}^{*\infty}\right.\right)}{\int_{y^{n+1}}\cdots\int_{y^{\infty}}\mathcal{L}\left(\boldsymbol{y}^{\infty}\left|\boldsymbol{\theta}^{*\infty}\right.\right)d\boldsymbol{y}^{n+1}\cdots d\boldsymbol{y}^{\infty}}
\]
and 
\[
\left.\mathbb{P}_{\infty}\pi^{\otimes\infty}\right|_{\boldsymbol{y}_{1{:}n}}=\frac{\int_{\mathbb{R}^{J\otimes\infty},\mathbb{R}_{+}}\mathcal{L}\left(\boldsymbol{y}^{\infty}\left|\boldsymbol{\theta}^{\infty}\right.\right)\pi\left(\boldsymbol{\theta}^{\infty}\right)d\boldsymbol{\theta}^{\infty}}{\int_{y^{n+1}}\cdots\int_{y^{\infty}}\int_{\mathbb{R}^{J\otimes\infty},\mathbb{R}_{+}}\mathcal{L}\left(\boldsymbol{y}^{\infty}\left|\boldsymbol{\theta}^{\infty}\right.\right)\pi\left(\boldsymbol{\theta}^{\infty}\right)d\boldsymbol{\theta}^{\infty}d\boldsymbol{y}^{n+1}\cdots d\boldsymbol{y}^{\infty}}.
\]
Given this notation, to prove (\ref{eq:Consis}), we need to show,
\[
\sup_{A\in\mathbb{R}^{\infty}}\left|\left.\mathbb{P}_{\infty}^{*}\right|_{\boldsymbol{y}_{1{:}n}}-\left.\mathbb{P}_{\infty}\pi^{\otimes\infty}\right|_{\boldsymbol{y}_{1{:}n}}\right|\rightarrow0,\ \textrm{in }\mathbb{P}_{\infty}^{*},
\]
and to prove (\ref{eq:TotalConsis}), we need to show, 
\begin{equation}
\lim_{n\rightarrow\infty}\sup_{A\in\mathbb{R}^{\infty}}\left|\left.\mathbb{P}_{\infty}^{*}\right|_{\boldsymbol{y}_{1{:}n}}\otimes p^{*}\left(\boldsymbol{y}_{1{:}n}\right)-\left.\mathbb{P}_{\infty}\pi^{\otimes\infty}\right|_{\boldsymbol{y}_{1{:}n}}\otimes p^{*}\left(\boldsymbol{y}_{1{:}n}\right)\right|=0.\label{eq:Consis-2}
\end{equation}
Since neither is greater than 2, and given that the equality, 
\[
\sup_{A\in\mathbb{R}^{\infty}}\left|\left.\mathbb{P}_{\infty}^{*}\right|_{\boldsymbol{y}_{1{:}n}}\otimes p^{*}\left(\boldsymbol{y}_{1{:}n}\right)-\left.\mathbb{P}_{\infty}\pi^{\otimes\infty}\right|_{\boldsymbol{y}_{1{:}n}}\otimes p^{*}\left(\boldsymbol{y}_{1{:}n}\right)\right|=\int_{y^{n}}\sup_{A\in\mathbb{R}^{\infty}}\left|\left.\mathbb{P}_{\infty}^{*}\right|_{\boldsymbol{y}_{1{:}n}}-\left.\mathbb{P}_{\infty}\pi^{\otimes\infty}\right|_{\boldsymbol{y}_{1{:}n}}\right|p^{*}\left(\boldsymbol{y}_{1{:}n}\right)d\boldsymbol{y}_{1{:}n},
\]
holds, they are equivalent. Thus, proving (\ref{eq:Consis-2}) is
sufficient to prove both.

Now, since we assumed $0<\pi\left(\mu^{*},\left\{ \boldsymbol{\beta}_{j}^{*}\right\} _{j=0,\cdots,J}\right)$,
$\mathbb{P}_{\infty}^{*}$ is absolute continuous with regard to the
marginal, $\mathbb{P}_{\infty}\pi^{\otimes\infty}$: $\mathbb{P}_{\infty}^{*}\ll\mathbb{P}_{\infty}\pi^{\otimes\infty}$.
Denote the likelihood ratio of $\mathbb{P}_{\infty}^{*}$ and $\mathbb{P}_{\infty}\pi^{\otimes\infty}$,
and its $\mathcal{F}_{n}$-conditional likelihood ratio as 
\[
Z=\frac{d\mathbb{P}_{\infty}^{*}}{d\mathbb{P}_{\infty}\pi^{\otimes\infty}},\quad Z_{n}=\mathbb{E}^{\mathbb{P}_{\infty}\pi^{\otimes\infty}}\left[\left.\frac{d\mathbb{P}_{\infty}^{*}}{d\mathbb{P}_{\infty}\pi^{\otimes\infty}}\right|\mathcal{F}_{n}\right].
\]
Here, $\mathbb{E}^{\mathbb{P}_{\infty}\pi^{\otimes\infty}}\left[X\right]$
is an integral regarding the cylinder measure, $\mathbb{P}_{\infty}\pi^{\otimes\infty}$.
Further, since $\mathbb{E}^{\mathbb{P}_{\infty}\pi^{\otimes\infty}}\left[Z_{n}\right]\leq1,\ ^{\forall}n$,
and from Doob's martingale convergence theorem, we have 
\begin{equation}
\lim_{n\rightarrow\infty}\mathbb{E}^{\mathbb{P}_{\infty}\pi^{\otimes\infty}}\left[\left|Z_{n}-Z\right|\right]=0.\label{eq:DMCT}
\end{equation}
From this, we can prove (\ref{eq:Consis-2}). For the function, $h$,
on $\mathbb{R}^{\infty}$, we have 
\begin{alignat*}{1}
 & \sup_{\left\Vert h\left(X,y^{n}\right)\right\Vert \leq1}\left|\int_{y^{n}}\left\{ \int h\left(\boldsymbol{y}^{n+1:\infty},\boldsymbol{y}_{1{:}n}\right)\left(\left.\mathbb{P}_{\infty}^{*}\right|_{\boldsymbol{y}_{1{:}n}}-\left.\mathbb{P}_{\infty}\pi^{\otimes\infty}\right|_{\boldsymbol{y}_{1{:}n}}\right)d\boldsymbol{y}^{n+1:\infty}\right\} p^{*}\left(\boldsymbol{y}_{1{:}n}\right)d\boldsymbol{y}_{1{:}n}\right|\\
= & \sup_{\left\Vert h\left(X,y^{n}\right)\right\Vert \leq1}\left|\int_{y^{n}}\int h\left(\boldsymbol{y}^{n+1:\infty},\boldsymbol{y}_{1{:}n}\right)\left\{ Z-Z_{n}\right\} \left.\mathbb{P}_{\infty}\pi^{\otimes\infty}\right|_{\boldsymbol{y}_{1{:}n}}d\boldsymbol{y}^{n+1:\infty}\mathbb{P}_{n}\pi^{\otimes n}d\boldsymbol{y}_{1{:}n}\right|\\
\leq & \mathbb{E}^{\mathbb{P}_{\infty}\pi^{\otimes\infty}}\left[\left|Z_{n}-Z\right|\right]\rightarrow0.
\end{alignat*}
If we let $h\left(X,y^{n}\right)=1_{A}\left(X\right)$, we obtain
the result. 
\end{proof}

\section{Step-by-step MCMC Algorithm of BCS }
\label{sec:NNGP-pos}

The use of the $m$-nearest neighbor Gaussian process for $\beta_j(X)$ leads to a multivariate normal distribution with a sparse precision matrix for $(\beta_j(x_1),\ldots,\beta_j(x_n))$, defined as 
$$
\pi(\beta_j(x_1),\ldots,\beta_j(x_n))
=\prod_{i=1}^n \phi(\beta_j^{\ast}(x_i);B_{\beta_j}(x_i)\beta_j^{\ast}(N(x_i)), \tau_{\beta_j}^2 F_{\beta_j}(x_i)), \ \ \ \ j=0,\ldots,J,
$$
with $\beta_j^{\ast}(x_i)=\beta_j(x_i)-\barbeta_j$, where 
\begin{equation}\label{eq:BF-beta}
\begin{split}
B_{\beta_j}(x_i)&=C_{\beta_j}(x_i, N(x_i); \phi_{\beta_j})C_{\beta_j}(N(x_i), N(x_i); \phi_{\beta_j})^{-1}, \\ 
F_{\beta_j}(x_i)&=1-C_{\beta_j}(x_i, N(x_i); \phi_{\beta_j})C_{\beta_j}(N(x_i), N(x_i); \phi_{\beta_j})^{-1}C_{\beta_j}(N(x_i), x_i; \phi_{\beta_j}),
\end{split}
\end{equation}
and $N(x_i)$ denotes an index set of $m$-nearest neighbors of $x_i$.
Here $C_{\beta_j}(t, v; \phi_{\beta_j})$ for $t=(t_1,\ldots,t_{d_t})\in \mathbb{R}^{d_t}$ and $v=(v_1,\ldots,v_{d_v})\in \mathbb{R}^{d_v}$ denotes a $d_t\times d_v$ correlation matrix whose $(i_t, i_v)$-element is $C(\|Z_{i_t}-v_{i_v}\|; \phi_{\beta_j})$.
In the same way, the joint prior distribution of $\mu$ is 
$$
\pi(\mu(z_1),\ldots,\mu(z_n))
=\prod_{i=1}^n \phi(\mu^{\ast}(z_i);B_{\mu}(z_i)\mu^{\ast}(N(z_i)), \tau_{\mu}^2 F_{\mu}(z_i)), 
$$
with $\mu^{\ast}(x_i)=\mu(x_i)-\barmu$, where 
\begin{equation}\label{eq:BF-mu}
\begin{split}
B_{\mu}(z_i)&=C_{\mu}(z_i, N(z_i);\phi_\mu)C_{\mu}(N(z_i), N(z_i);\phi_\mu)^{-1}, \\ 
F_{\mu}(z_i)&=1-C_{\mu}(z_i, N(z_i);\phi_\mu)C_{\mu}(N(z_i), N(z_i);\phi_\mu)^{-1}C_{\mu}(N(z_i), z_i;\phi_\mu),
\end{split}
\end{equation}
and $C_\mu(X,X;\phi_\mu)$ is defined in the same way as $C_{\beta_j}(X,X;\phi_{\beta_j})$. 
In what follows, we assume that $f_{ji}\sim N(a_{ji}, b_{ji})$, where $a_{ji}\equiv a_j(x_i)$ and $b_{ji}\equiv b_j(x_i)$.

\bigskip
The full conditional distributions of $\Phi_n$ and $\psi$ are described as follows: 
\begin{itemize}
\item[-]
{\bf (Sampling of varying coefficients)} \ \ For $i=1,\ldots,n$, the full conditional distribution of $(\beta_0(x_i),\ldots,\beta_J(x_i))$ is given by $N(\barbeta_j + A_{i}^{-1}B_{i}, A_{i}^{-1})$ with
\begin{align*}
&A_{i}=\frac{Z_if_if_i^{\top}}{\sigma^2}+{\rm diag}(\gamma_{0i},\ldots,\gamma_{Ji}), \ \ \ \  
B_{i}=\frac{Z_if_i(y_i-\mu_i)}{\sigma^2}+(m_{0i},\ldots,m_{Ji})^\top,
\end{align*}
where 
\begin{align*}
&\gamma_{ji}=\frac{1}{\tau_{\beta_j}^2 F_{\beta_j}(x_i)}+\sum_{t;x_i\in N(t) }\frac{B_{\beta_j}(t; x_i)^2}{\tau_{\beta_j}^2 F_{\beta_j}(t)}, \\
&m_{ji}=\frac{B_{\beta_j}(x_i)^\top \beta_j^{\ast}(N(x_i))}{\tau_{\beta_j}^2F_{\beta_j}(x_i)} +\sum_{t;x_i\in N(t) }\frac{B_{\beta_j}(t;x_i)}{\tau_{\beta_j}^2F_{\beta_j}(t)}\Big\{\beta_j^{\ast}(t)-\sum_{s\in N(t), s\neq x_i}B_{\beta_j}(t;s) \beta_j^{\ast}(s)\Big\},
\end{align*}
and $f_i=(1, f_{1i},\ldots,f_{Ji})^\top$.
Here $B_{\beta_j}(t;s)$ denotes the scalar coefficient for $\beta_j^{\ast}(x_i)$ among the element of the coefficient vector $B_{\beta_j}(t)$.

\item[-]
{\bf (Sampling of prognostic term)} \ \ For $i=1,\ldots,n$, the full conditional distribution of $\mu(z_i)$ is given by 
$$
N\left(
\barmu + \left(\frac1{\sigma^2}+\gamma_i^{(\mu)}\right)^{-1}\left(\frac{\widetilde{y}_i}{\sigma^2}+m_i^{(\mu)}\right), \left(\frac1{\sigma^2}+\gamma_i^{(\mu)}\right)^{-1}
\right),
$$
where $\widetilde{y}_i=y_i-Z_i\{\beta_0(x_i)+\sum_{j=1}^J \beta_j(x_i)f_j(x_i)\}$, and 
\begin{align*}
&\gamma_{i}^{(\mu)}=\frac{1}{\tau_\mu^2 F_{\mu}(x_i)}+\sum_{t;z_i\in N(t)}\frac{B_{\mu}(t; z_i)^2}{\tau_{\mu}^2 F_{\mu}(t)}, \\
&m_{i}^{(\mu)}=\frac{B_{\mu}(z_i)^\top \mu^{\ast}(N(z_i))}{\tau_{\mu}^2F_{\mu}(z_i)} +\sum_{t;z_i\in N(t) }\frac{B_{\mu}(t;z_i)}{\tau_{\mu}^2F_{\mu}(t)}\Big\{\mu^{\ast}(t)-\sum_{s\in N(t), s\neq z_i}B_{\mu}(t;s) \mu^{\ast}(s)\Big\},
\end{align*}

\item[-]
{\bf (Sampling of latent factors)}\ \ \ 
Generate $f_{ji}$ from $N((A_{ji}^{(f)})^{-1}B_{ji}^{(f)}, (A_{ji}^{(f)})^{-1})$, where 
$$
A_{ji}^{(f)}=\bigg(\frac{\beta_{ji}^2}{\sigma^2}+\frac1{b_{ji}}\bigg), \ \ \ \ \ 
B_{ji}^{(f)}=\frac{\beta_{ji}}{\sigma^2}\bigg\{y_i-\mu_i-Z_i\bigg(\beta_{0i}-\sum_{k\neq j}\beta_{ki}f_{ki}\bigg)\bigg\}+\frac{a_{ji}}{b_{ji}}
$$

\item[-]
{\bf (Sampling of $\tau_{\beta_j}^2$)}\ \ 
For $j=0,\ldots,J$, the full conditional distribution of $\tau_{\beta_j}^2$ is 
$$
{\rm IG}\left(\frac{\delta_{\beta_j}+n}{2}, \frac{\eta_{\beta_j}}{2}+\frac12\sum_{i=1}^n \frac{\big\{\beta_j^{\ast}(x_i)-B_{\beta_j}(x_i)\beta_j^{\ast}(N(x_i))\big\}^2}{F_{\beta_j}(x_i)}\right).
$$

\item[-]
{\bf (Sampling of $\tau_{\mu}^2$)}\ \ 
The full conditional distribution of $\tau_{\mu}^2$ is 
$$
{\rm IG}\left(\frac{\delta_\mu+n}{2}, \frac{\eta_{\mu}}{2}+\frac12\sum_{i=1}^n \frac{\big\{\mu^{\ast}(z_i)-B_{\mu}(z_i)\mu^{\ast}(N(z_i))\big\}^2}{F_{\mu}(z_i)}\right).
$$

\item[-]
{\bf (Sampling of $\phi_{\beta_j}$)} \ \ 
For $j=0,\ldots,J$, the full conditional distribution of $\phi_{\beta_j}$ is proportional to 
$$
\prod_{i=1}^n \phi(\beta_j^{\ast}(x_i);B_{\beta_j}(x_i)\beta_j^{\ast}(N(x_i)), \tau_{\beta_j}^2F_{\beta_j}(x_i)), \ \ \  \phi_{\beta_j}\in (\underline{c}_{\beta}, \overline{c}_{\beta}),
$$
where $B_{\beta_j}(x_i)$ and $F_{\beta_j}(x_i)$ depend on $\phi_{\beta_j}$ as defined in (\ref{eq:BF-beta}).

\item[-]
{\bf (Sampling of $\phi_{\mu}$)} \ \ 
The full conditional distribution of $\phi_{\mu}$ is proportional to 
$$
\prod_{i=1}^n \phi(\mu^{\ast}(z_i);B_{\mu}(z_i)\mu^{\ast}(N(z_i)), \tau_{\mu}^2F_{\mu}(z_i)), \ \ \  \phi_{\mu}\in (\underline{c}_{\mu}, \overline{c}_{\mu}),
$$
where $B_{\mu}(z_i)$ and $F_{\mu}(z_i)$ depend on $\phi_{\mu}$ as defined in (\ref{eq:BF-mu}).

\item[-] 
{\bf (Sampling of $\sigma^2$)} \ \ The full conditional is $\sigma^2\sim {\rm IG}(\delta_\sigma/2+n/2,\eta_{\sigma}/2+\sum_{i=1}^n(\widetilde{y}_i-\mu_i)^2/2)$, where $\widetilde{y}_i=y_i-Z_i\{\beta_0(x_i)+\sum_{j=1}^J \beta_j(x_i)f_j(x_i)\}$.
\end{itemize}

\section{Additional Simulation Studies }
\label{sec:supp-sim}

\subsection{Experiments under RCT}

We here investigate the performance of BCS and existing methods under randomized control trials (RCT).
We adopt the same data generating process as in Section~\ref{sec:sim-syn}, except that there is no unobserved confounders, described as 
$$
Y=\mu(X)+\tau(X)T+\ep,  \ \ \ \ \ep\sim N(0, \sigma^2),
$$
where $T\sim {\rm Ber}(0.5)$ is the randomized treatment assignment and $\sigma^2=1$, and the scenarios of $\mu(X)$ and $\tau(X)$ \citep[taken from][]{hahn2020bayesian} are given by
\begin{align*}
&\text{(A)} \ \ \mu(X)=-7+6|X_3|-3X_5, \ \ \ \ \ \text{(B)} \ \ \mu(X)=2+2\sin(3X_3)\\
&\text{(A)} \ \ \tau(X)=1+2X_2X_5, \ \ \ \ \ \text{(B)} \ \ \tau(X)=1+2X_2X_5+X_3^2/2.
\end{align*}
We adopt four scenarios consisting of all the possible combinations. 
For each scenario, we considered three scenarios of the sample size, $n$ and the number of covariates, $p$, as $(n,p)=(50, 5), (300, 20)$ and $(300, 60)$. 
We generate random samples from the data generating process and estimate the form of $\tau(X)$ to evaluate the treatment effects of the observed samples.

For the same methods in Section~\ref{sec:sim-syn}, we evaluated RMSE of point estimates and CP and AL of $95\%$ credible intervals, where the results are shown in Tables~\ref{tab:sim-mse-supp} and \ref{tab:sim-cp-supp}.
It is observed that MSE of BCS is smallest or second smallest among the methods in all the scenarios, showing its notable performance in RCT settings. 
Furthermore, Tables~\ref{tab:sim-cp-supp} shows that existing methods fail to produce valid credible intervals and the coverage probability is far below from the nominal level, especially when the number of covariates is large. 
On the other hand, BCS can successfully evaluate the uncertainty of HTE estimation such that the coverage probability of $95\%$ credible intervals are around nominal levels in all the scenarios.

\begin{table}[t!]
\caption{Mean squared error (MSE) of point estimates of heterogeneous treatment effects, averaged over 100 Monte Carlo replications.
The smallest and second smallest MSE values are highlighted in bold. 
}
\label{tab:sim-mse-supp}
\begin{center}
\begin{tabular}{ccccccccccccc}
\hline
$\mu$ & $\tau$ & $(n,p)$ &  & BCS & BCF & LM & AM & CF & XL & RL & CST \\
\hline
 &  & (50, 5) &  & {\bf 4.81} & 10.55 & 8.42 & {\bf 4.06} & 7.36 & 11.37 & 19.74 & 9.51 \\
(A) & (A) & (300, 20) &  & {\bf 2.01} & 3.40 & 3.70 & 2.91 & {\bf 2.47} & 4.59 & 4.23 & 4.25 \\
 &  & (300, 60) &  & {\bf 2.32} & 3.82 & 4.06 & {\bf 2.86} & 2.99 & 5.97 & 7.59 & 5.00 \\
 \hline
 &  & (50, 5) &  & {\bf 3.21} & 7.28 & 4.69 & {\bf 4.46} & 6.06 & 10.04 & 19.22 & 9.51 \\
(B) & (A) & (300, 20) &  & {\bf 1.94} & 2.58 & 3.01 & 2.97 & {\bf 2.26} & 3.82 & 3.63 & 3.59 \\
 &  & (300, 60) &  & {\bf 2.24} & 3.09 & 3.11 & 2.98 & {\bf 2.62} & 4.80 & 6.08 & 5.26 \\
 \hline
 &  & (50, 5) &  & {\bf 5.01} & 11.38 & 9.48 & {\bf 4.25} & 8.30 & 12.09 & 19.79 & 11.26 \\
(A) & (B) & (300, 20) &  & {\bf 2.22} & 4.00 & 4.48 & 2.95 & {\bf 2.93} & 5.25 & 4.88 & 5.06 \\
 &  & (300, 60) &  & {\bf 2.63} & 4.32 & 5.05 & {\bf 2.99} & 3.42 & 6.12 & 8.14 & 6.18 \\
 \hline
 &  & (50, 5) &  & {\bf 3.60} & 7.91 & 5.38 & {\bf 5.06} & 6.82 & 9.93 & 19.56 & 9.59 \\
(B) & (B) & (300, 20) &  & {\bf 2.08} & 2.84 & 3.40 & 2.99 & {\bf 2.41} & 4.38 & 4.10 & 3.84 \\
 &  & (300, 60) &  & {\bf 2.46} & 3.59 & 3.61 & 3.09 & {\bf 3.00} & 5.46 & 6.71 & 6.43 \\
\hline
\end{tabular}
\end{center}
\end{table}

\begin{table}[t!]
\caption{CP ($\%$) and AL of $95\%$ credible/confidence intervals of heterogeneous treatment effects, averaged over 100 Monte Carlo replications.
}
\label{tab:sim-cp-supp}
\begin{center}
\begin{tabular}{ccccrrrrrrrrrrr}
\hline
         &       &        &    & \multicolumn{1}{c}{}    & \multicolumn{1}{c}{}    & \multicolumn{1}{c}{CP} & \multicolumn{1}{c}{}   & \multicolumn{1}{c}{}   & \multicolumn{1}{c}{} & \multicolumn{1}{c}{}    & \multicolumn{1}{c}{}    & \multicolumn{1}{c}{AL} & \multicolumn{1}{c}{}   & \multicolumn{1}{c}{}   \\
$\mu$ & $\tau$ & $(n,p)$ &  & BCS & BCF & LM & AM & CF &  & BCS & BCF & LM & AM & CF \\
\hline
 &  & (50, 5) &  & 96.0 & 85.3 & 84.3 & 68.3 & 48.5 &  & 9.07 & 7.93 & 7.78 & 3.68 & 5.02 \\
(A) & (A) & (300, 20) &  & 98.2 & 85.4 & 54.0 & 49.6 & 83.9 &  & 6.09 & 4.61 & 2.54 & 2.02 & 4.90 \\
 &  & (300, 60) &  & 98.2 & 83.9 & 51.7 & 50.6 & 70.8 &  & 6.40 & 4.66 & 2.70 & 2.08 & 4.33 \\
 \hline
 &  & (50, 5) &  & 94.6 & 82.8 & 85.2 & 69.6 & 38.4 &  & 7.02 & 5.93 & 5.92 & 4.06 & 3.54 \\
(B) & (A) & (300, 20) &  & 95.3 & 89.0 & 46.5 & 45.0 & 82.5 &  & 5.02 & 4.31 & 1.67 & 1.89 & 4.24 \\
 &  & (300, 60) &  & 93.8 & 86.2 & 45.9 & 44.4 & 67.0 &  & 4.96 & 4.36 & 1.82 & 1.85 & 3.19 \\
 \hline
 &  & (50, 5) &  & 96.5 & 83.9 & 85.3 & 68.4 & 49.1 &  & 9.34 & 7.94 & 8.39 & 3.76 & 5.06 \\
(A) & (B) & (300, 20) &  & 98.1 & 83.7 & 55.7 & 52.0 & 81.9 &  & 6.38 & 4.77 & 3.01 & 2.15 & 5.14 \\
 &  & (300, 60) &  & 97.5 & 83.0 & 56.1 & 52.5 & 69.2 &  & 6.67 & 4.89 & 3.34 & 2.27 & 4.58 \\
 \hline
 &  & (50, 5) &  & 93.5 & 81.8 & 84.1 & 69.6 & 37.5 &  & 7.30 & 6.14 & 6.12 & 4.19 & 3.61 \\
(B) & (B) & (300, 20) &  & 94.1 & 87.5 & 42.7 & 49.0 & 77.1 &  & 5.10 & 4.45 & 1.78 & 2.05 & 4.03 \\
 &  & (300, 60) &  & 92.0 & 84.0 & 42.6 & 48.0 & 63.1 &  & 5.09 & 4.50 & 1.93 & 2.06 & 3.13 \\
\hline
\end{tabular}
\end{center}
\end{table}

\subsection{Out-of-sample performance}
We further investigate the out-of-samples performance of estimation and inference on heterogeneous treatment effects. 
Again, adopting the same data generating scenarios as in Section~\ref{sec:sim}, we apply the same methods to the observed data to estimate heterogeneous treatment effects in 200 test samples generated in the same way as the observed samples.
Regarding the proposed BCS, we combine the results from CF, LM and AM. 
Mean squared error (MSE) of point estimates and coverage probability (CP) and average length (AL) of $95\%$ credible/confidence intervals of heterogeneous treatment effects, averaged over 100 replications, are reported in Tables~\ref{tab:out-of-sample-MSE} and \ref{tab:out-of-sample-coverage}.
As confirmed in Section~\ref{sec:sim-syn}, the proposed BCS attains the smallest and second smallest MSE values in all of the scenarios, and it significantly improves the performance of the three models  synthesized. 
Table~\ref{tab:out-of-sample-coverage} shows that BCS provides reasonable interval estimation, where the empirical coverage is around the nominal level, while the other methods severely undercover the true value.

\begin{table}[htb!]
\caption{Mean squared error (MSE) of point estimates of heterogeneous treatment effect in 200 test samples, averaged over 100 replications.  
The smallest and second smallest MSE values are highlighted in bold.  
}
\label{tab:out-of-sample-MSE}
\begin{center}
\begin{tabular}{ccccccccccccccccc}
\hline
$\mu$ & $\tau$ & $(n,p)$ &  & BCS & BCF & LM & AM & CF & XL & RL & CST \\
\hline
 &  & (50, 5) &  & {\bf 5.47} & 12.45 & 9.15 & {\bf 4.77} & 20.22 & 9.61 & 12.58 & 10.67 \\
(A) & (A) & (300, 20) &  & {\bf 3.38} & 4.00 & 5.71 & 3.62 & 4.77 & {\bf 2.99} & 4.51 & 4.48 \\
 &  & (300, 60) &  & 4.34 & 4.34 & 13.89 & {\bf 3.77} & 8.24 & {\bf 3.56} & 5.20 & 6.36 \\
 \hline
 &  & (50, 5) &  & {\bf 4.92} & 8.67 & 6.56 & {\bf 5.99} & 19.82 & 8.01 & 11.07 & 10.70 \\
(B) & (B) & (300, 20) &  & {\bf 2.70} & 3.56 & 4.80 & 3.59 & 4.64 & {\bf 3.15} & 4.37 & 4.77 \\
 &  & (300, 60) &  & {\bf 3.60} & 4.10 & 9.57 & {\bf 3.72} & 7.07 & 3.77 & 4.98 & 6.36\\
\hline
\end{tabular}
\end{center}
\end{table}

\begin{table}[htb!]
\caption{Coverage probability (CP) and average length (AL) of $95\%$ credible/confidence intervals of heterogeneous treatment effect in 200 test samples, averaged over 100 replications.  
}
\label{tab:out-of-sample-coverage}
\begin{center}
\begin{tabular}{ccccccccccccccccc}
\hline
&&&&  \multicolumn{5}{c}{CP ($\%$)} && \multicolumn{5}{c}{AL}\\
$\mu$ & $\tau$ & $(n,p)$ &  & BCS & BCF & LM & AM & CF &  & BCS & BCF & LM & AM & CF \\
\hline
 &  & (50, 5) &  & 96.7 & 81.9 & 85.0 & 67.4 & 48.8 &  & 10.28 & 7.85 & 8.32 & 3.86 & 5.00 \\
(A) & (A) & (300, 20) &  & 99.3 & 83.8 & 83.3 & 47.2 & 81.7 &  & 8.69 & 4.72 & 6.41 & 2.09 & 4.91 \\
 &  & (300, 60) &  & 99.4 & 83.3 & 92.0 & 48.6 & 69.2 &  & 10.84 & 4.82 & 13.01 & 2.25 & 4.15 \\
 \hline
 &  & (50, 5) &  & 95.3 & 81.8 & 81.7 & 66.7 & 36.9 &  & 9.34 & 6.41 & 6.39 & 4.31 & 3.56 \\
(B) & (B) & (300, 20) &  & 97.8 & 86.0 & 74.9 & 44.0 & 74.4 &  & 7.20 & 4.68 & 4.74 & 1.92 & 4.06 \\
 &  & (300, 60) &  & 97.2 & 83.1 & 88.9 & 45.1 & 61.1 &  & 8.06 & 4.76 & 9.86 & 2.04 & 3.07 \\

\hline
\end{tabular}
\end{center}
\end{table}

\end{document}